\documentclass[preprint,times,authoryear]{elsarticle}

\usepackage{amsmath}
\usepackage{mathtools}
\usepackage[T1]{fontenc}
\usepackage{graphicx}
\usepackage{float}
\usepackage{color}
\usepackage{caption}
\usepackage{epstopdf}
\usepackage{lettrine}
\usepackage{fancyhdr}
\usepackage{anysize}
\usepackage{comment}
\usepackage{subfigure}
\usepackage{wrapfig} 
\usepackage{multicol} 
\usepackage[percent]{overpic} 
\usepackage{wasysym} 
\usepackage{cancel}  
\usepackage{mathenv} 
\usepackage{enumerate} 
\usepackage[toc,page]{appendix}
\usepackage{empheq} 
\usepackage{units} 
\usepackage{multirow}

\usepackage[colorlinks,citecolor=red,urlcolor=blue,bookmarks=false,hypertexnames=true]{hyperref}

\begin{document}

\begin{frontmatter}

\title{Subsistence of ice-covered lakes during the Hesperian at Gale crater, Mars}

\author[add1,add2]{Alexandre M. Kling}
\author[add2]{Robert M. Haberle}
\author[add3]{Christopher P. McKay}
\author[add4]{Thomas F. Bristow}
\author[add5]{Frances\;Rivera-Hernandez}

\address[add1]{Bay Area Environmental Research Institute, Moffett Field, CA 94035}
\address[add2]{Planetary Systems Branch, NASA Ames Research Center, Moffett Field, CA 94035}
\address[add3]{Space Science and Astrobiology Division, NASA Ames Research Center, Moffett Field, CA 94035, USA}
\address[add4]{Exobiology Branch, NASA Ames Research Center, Moffett Field, CA 94035}
\address[add5]{Dartmouth College, Davis, Hanover, NH 03755}

\begin{abstract}

\indent
Sedimentary deposits characterized by the Mars Science Laboratory \textit{Curiosity} rover provide evidence that Gale crater, Mars  intermittently hosted a fluvio-lacustrine environment during the Hesperian. However, estimates of the CO$_2$ content of the atmosphere at the time the sediments in Gale crater were deposited are far less than needed by any climate model to maintain temperatures warm enough for sustained open water lake conditions due to the low solar energy input available at that time. To reconcile some of the in-situ sedimentological evidence for liquid water with climate modeling studies, we perform the water budget of evaporation against precipitation to estimate the minimum lifetimes and the rainfall requirements for open water and ice-covered lakes in Gale crater, for a wide range of pressures and temperatures.  We found that both open water and ice-covered lakes are possible, and that ice-covered lakes provide better consistency in regards of  the low erosion rates estimates for the Hesperian.  We  incrementally test the existence of open water conditions using energy balance calculations for the global, regional, and seasonal temperatures, and we assess if the preservation of liquid water was possible under perennial ice covers.   We found scenarios where lacustrine conditions are preserved in a cold climate, where the resupply of water by the inflow of rivers and high precipitation rates are substituted by an abutting glacier. For equatorial temperatures as low as 240K-255K, the ice thickness ranges from 3-10 m, a value  comparable to the range of those for the perennially ice-covered lakes in Antarctica (3-6 m). The ice-covered lake hypothesis is a compelling way to decouple the mineralogy and the climate by limiting the gas exchanges between the sediment and the CO$_2$  atmosphere, and it eliminates the requirement for global mean temperatures above the freezing point. Not only do ice-covered lakes provide a baseline for exploring the range of possible lake scenarios for Gale crater during the Hesperian that is fully consistent with climate studies, but also they might have been ideal environments to sustain life on Mars.

\end{abstract}

\end{frontmatter}

\tableofcontents

\newpage

\section{Introduction}

\begin{figure*}
\centering
\includegraphics[width=\textwidth, trim=0cm 0 0cm 0 ]{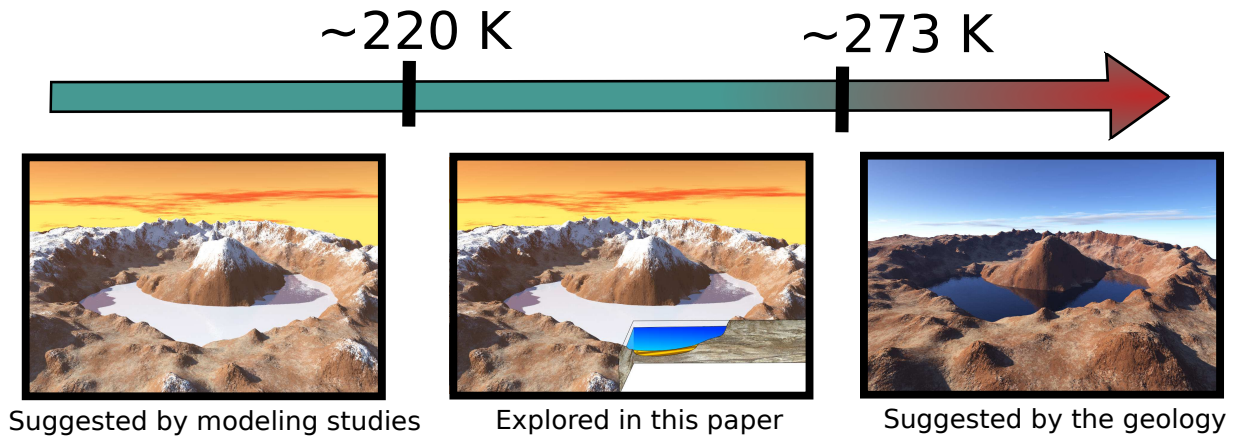} \\
\caption{Three climatic scenarios for Gale crater Mars, during the Hesperian}
\label{mars_landscape_fig}
\end{figure*}

Detailed observations of the sedimentary textures (e.g. grain size and shape) and structures (e.g. lamination and cross-stratification) of rocks encountered by The Mars Science Laboratory (MSL) \textit{Curiosity} rover  has revealed geological evidence for ancient lake and river depositions in Gale crater, Mars, $\sim$3.6 Gya to 3.2 Gya  \citep{grotzinger_2014,grotzinger_2015,edgar_2017}.  Although liquid water subsists in the perennially ice-covered lakes in Antarctica where the mean annual temperatures stay below freezing (e.g. \citet{doran_1994}; \citet{sokratova_2011}), the absence of glaciogenic sedimentary deposits in Gale crater has helped to support the idea of an early "wet and warm" Mars \citep{grotzinger_2015}.  Based on crater counts, Gale crater's age is at the Noachian/ Hesperian transition \citep{le_deit_2013,thomson_2011}, and the sediment inside the crater is thought to have been deposited over a few hundred million years, well into the Hesperian \citep{palucis_2014,thomson_2011,grant_2014,grotzinger_2015}, a period where modeling studies have consistently predicted a cold climate (see \citet{forget_2017} and \citet{haberle_2017} for a review of early Mars climate studies). 

By analogy with lakes on Earth, there are  two main requirements for the subsistence of a liquid lake on Mars that will be investigated in this study: the first is driven by the water budget, meaning that the rain or snow  or precipitations into the lake's basin must balance the evaporative losses (other sources like groundwater as in \citet{tosca_2018} will be omitted in order to place an upper limit on the precipitation rates).  This dictates whether or not a body of water can exist in the first place, either frozen (Figure \ref{mars_landscape_fig}, left), or as open water (Figure \ref{mars_landscape_fig}, right). The second requirement relates to whether the local surface temperatures are annually or seasonally above the freezing point, which is taken as 273K throughout the paper. The presence of perchlorate salts has been identified at numerous sites on Mars (including at Gale crater \citet{martin-torres_2015}) and has the potential of lowering the freezing  point of water by tens of kelvin at high concentration \citep{mohlmann_2011}. However,  the actual composition and concentration of salts in the potential lake remain unconstrained. Pure water is used as a limiting case since the presence of salts would only make it easier to reconcile cold climates with evidence for open water lakes. We assess if the thermodynamics of the lake system can allow for the preservation of a liquid environment under a perennial ice cover (Figure \ref{mars_landscape_fig}, center) and estimate what the thickness of the ice may be under different climate scenarios. Finally, we discuss the implications of the ice-covered lake hypothesis for the geology of Gale crater observed by the \textit{Curiosity} rover.

\section{Water budget: precipitation requirements for open water and frozen lakes}

\subsection*{Physics of water on Mars}

On present-day Mars, liquid water is not stable: even though certain low land areas (including Gale crater) may episodically experience conditions above the triple point (611Pa, 273K), any liquid water will either quickly evaporate away because of the low partial pressure of water vapor in the atmosphere, or freeze solid if a source of energy (radiation) is not available to compensate for evaporative cooling \citep{ingersoll_1971,haberle_2001,chittenden_2008}.  Therefore, computing the evaporation rate is a convenient way to constrain the climate  required to sustain bodies of water on early Mars. Above a lake, two physical processes can contribute to the removal of the saturated water vapor from the air/water interface: free convection, which is driven by the buoyancy of water vapor into the denser CO$_2$ atmosphere, and forced convection which is a result of the winds advecting the water vapor away from the source.

On Earth, boundary layers above non-heated lakes are nearly neutrally stable, thus the evaporation is dominated by forced convection \citep{adams_1990}. On the other hand, on Mars, due to the low atmospheric pressure, free convection is significant and temperature-dependent. \cite{chittenden_2008} found experimentally that at low relative humidity ($\approx 1\%$) and at a temperature of $-15 ^oC$, the evaporation rate is $E= 0.68+0.025V [mm. {hr}^{-1}]$ with $V$ the wind speed in $m. s^{-1}$. However, they observed almost no dependence of the evaporation rate on the wind speed in high relative humidity atmospheres ($RH=$ 30-35\%).

On early Mars, the situation could have been in between these two cases.  A potential caveat is that as the wind speed increases, turbulent mixing homogenizes the CO$_2$/water vapor mixture above the lake which tends to shut down the transport by buoyancy of the saturated air at the very surface of the lake toward the 'conceptually dry' ambient air. To mitigate an over-estimation of the evaporation rate, we use the norm of the free and forced evaporative fluxes in place of their linear sum, so the total evaporation quickly converges toward either the purely free or the purely convective asymptotes in the case where one process dominates the other  \citep{adams_1990}.

We first consider that the lakes during the Hesperian were either open water or ice-covered and following \citet{ingersoll_1971}, \citet{adams_1990} and \citet{dundas_2010} we compute the sublimation and evaporation rates combining free and forced convection:
\begin{equation}
\begin{split}
E_{combined}= &\frac{1}{\rho_{H_2O}}\left(\underbrace{\left(K_M\right)^2}_{free}+\underbrace{\left(A \; u \right)^2}_{forced}\right)^\frac{1}{2} \\
\times & \underbrace{\frac{M_{H_20}}{R}\left(\frac{e^{surf}_{sat}}{T_{surf}} -RH\frac{e^{atm}_{sat}}{T_{atm}} \right)}_{T \; and \; RH\;dependent} \quad [m/s]
\end{split}
\label{eq_combinedRH}
\end{equation}

with $\rho_{H_2O}$ the density of either water or ice which is used to convert the evaporation rate from $kg.m^{-2}s^{-1}$ to  $m.s^{-1}$, $K_M$ a bulk coefficient for the diffusion of water vapor into the denser CO$_2$ atmosphere in unit of m s$^{-1}$, $A$ a function characterizing the magnitude of the wind-driven  convection, $u$ the wind speed, $ M_{H_20}$ the molar mass of the  water, $R$ the gas constant, $RH$ the relative humidity, $T_{surf}$ and $T_{atm}$ respectively the lake's surface temperature and the atmospheric temperature (with $T_{surf} \geqslant T_{atm}$), $e^{surf}_{sat}$ and $e^{atm}_{sat}$ the saturated partial pressure of H$_2$O evaluated respectively at $T_{surf}$ and $T_{atm}$ over water or ice. The expressions used for $K_M$ and $A$ are given in the Appendix.

To assess which combinations of pressure and temperature could have been compatible with long lived lakes (>10,000 years e.g. \citet{grotzinger_2015}), we derive two metrics from the evaporation rate: a) the minimum lake lifetime and b) the precipitation rate required to sustain the constant water levels. Plausible values for the humidity, surface wind speed, geometry of the lake and water run-off are proposed in Table \ref{hesperian_table} and the derivation of the parameters from Eq. (\ref{eq_combinedRH}) is detailed in the  Appendix.

\subsection*{Minimum lake lifetime and precipitation rates}

First, the minimum lifetime of a lake is  obtained assuming no resupply of the lake through rainfall, ground water flow, or run-off. In other words the lifetime of a lake simply corresponds to the time required to evaporate the water column at the deepest point of the lake:

\begin{equation}
\tau_{lake}=\frac{h_{lake}}{E_{combined}}  \quad [yr]
\label{eq_lake_lifetime}
\end{equation}
with $h_{lake}$ the maximum depth of the lake, $E_{combined}$ the evaporation rate from Eq. (\ref{eq_combinedRH}) in unit of Earth years.

In order to compute the precipitation rate, we need to know the total catchment area. We approximate Gale crater as a parabolic crater of diameter $2R_{Gale}$=150 km, maximum depth $h_{Gale}=3km$  and without a central mound (the presence of a mound has no impact on the catchment area calculation). This geometry provides a reasonable fit of the present-day crater's topography, including the North/South asymmetry. The lake is represented  as a spherical cap to relate the radius of the lake to its maximum depth (see Figure \ref{spherical_cap_fig}):

\begin{figure}
\includegraphics[width=0.4\textwidth]{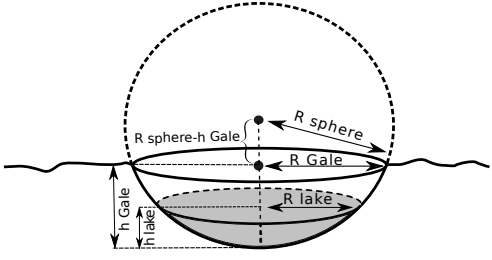}
 \caption{The spherical cap approximation for a lake (in grey) sitting at the bottom of a parabolic crater relates the depth of the lake to its catchment area}
 \label{spherical_cap_fig}
\end{figure}

\begin{equation}
 \begin{split}
&\begin{cases}
\left(R_{sphere}-h_{Gale}\right)^2+R_{Gale}^2=R_{sphere}^2  \hfill\\
\left(R_{sphere}-h_{lake}\right)^2+R_{lake}^2=R_{sphere}^2
\end{cases}\\
\Rightarrow &\begin{cases}
R_{sphere}=\dfrac{R_{Gale}^2+h_{Gale}^2}{2 h_{Gale}}  \hfill\\
R_{lake}=\sqrt{2 R_{sphere}\; h_{lake}-h_{lake}^2}
\end{cases}
 \end{split}
 \label{eq_spherical_cap}
\end{equation}

A 100m deep lake sitting in a 150km diameter/3 km deep parabolic crater has a diameter of approximately 27km, resulting in a catchment area $S_{catch}=\pi( R^2_{Gale}-R^2_{lake}) \approx$ 17000 km$^2$ about 29 times bigger than the lake's area itself. The contribution of the catchment areas up-slope of Peace Vallis (730 k$m^2$), Dulce Vallis  (450 k$m^2$) \citep{palucis_2014}, and other minor fans extending beyond the crater's rims make altogether less than 7 \% of the crater's catchment area and are neglected in this calculation.

The precipitation rates required to sustain the water level must exactly compensate the evaporation rates in a closed basin (i.e. no ground water recharge) \citep{matsubara_2011}:
\begin{equation}
\left(S_{lake}+ \alpha_{R} S_{catch}\right) P-S_{lake}E=0
\label{eq_lake_balance}
\end{equation}
with $S_{lake}$ the lake area, $\alpha_{R}$ the run-off coefficient which represents the fraction of the water deposited on the catchment area that enters the lake, $S_{catch}$ the catchment area, $P$ the precipitation rate and $E$ the evaporation rate. We assume that the precipitation that does not run-off either evaporates on the slopes or infiltrates a water table that is not connected to the lake system. The precipitation rate therefore represents a maximum rate since groundwater could have also fed the lake.

\begin{table*}
\footnotesize
\centering
\begin{tabular}{|c|c|c|}
\hline
Parameter &  Value & Justification \\
\hline
\hline
Relative humidity  & 50\% & humid climate, comparable to the annual mean value on the American Pacific Coast. \citep{NOAA_CCD_2015} \\
\hline
Surface winds      & 5 m/s & arbitrary value, typical for Mars based on climate modeling  \\
\hline
Run-off coefficient & 20\% & typical for sandy soils with  of <7\% slope \citep{ASCE_1993}\\
\hline
Depth of the lake & $\sim$100m & based on 75 m stratigraphy traversed by MSL \citep{grotzinger_2015}\\
\hline
\end{tabular}
\caption{Assumptions for our model}
\label{hesperian_table}
\end{table*}

Reorganizing Eq. (\ref{eq_lake_balance}), we have for the precipitation:

 \begin{equation}
P=\frac{E}{1+\alpha_{R} \left(\frac{R^2_{Gale}}{R^2_{lake}}-1\right)} \approx \frac{E}{1+ 29 \;\alpha_{R} } \quad [m/s]
\label{eq_lake_balance2}
\end{equation}

\subsection*{Implications for the Hesperian climate}

\begin{figure*}
\centering
\includegraphics[width=\textwidth, clip ,trim= 1cm 0 4cm 0 ]{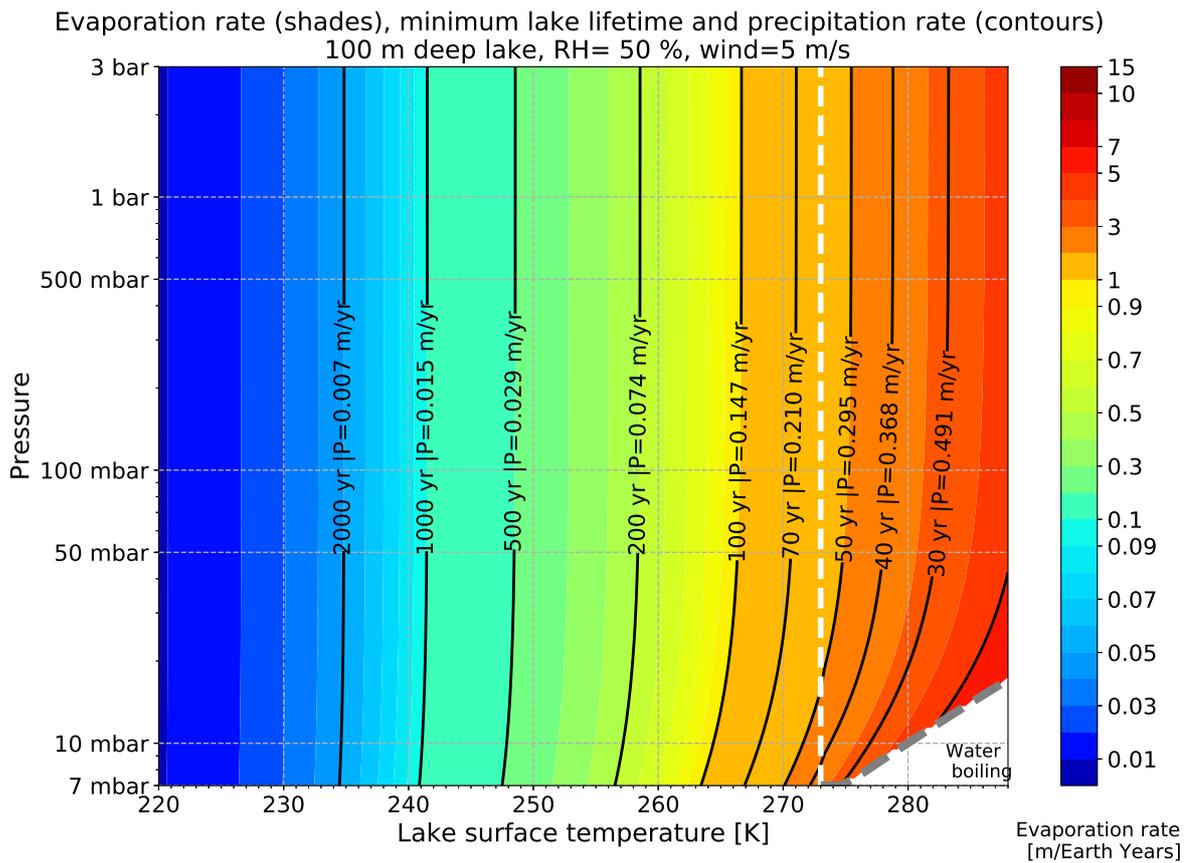} \\
\caption{Evaporation rate (shades), minimum lifetimes for a 100m deep lake and precipitation rates required to sustain the water level (black contours) as a function of the pressure and temperature. The white dashed line shows 273K.}
\label{lifetime_fig}
\end{figure*}

Using Eq. (\ref{eq_combinedRH}) and assuming $T_{surf}\approx T_{atm}=\overline{T_R}$, with $\overline{T_R}$ the annually-averaged regional temperature near Gale's latitude, we compute the evaporation rate, the minimum lifetime of a lake, and the precipitation rate required to sustain the water level, as a function of pressure and temperature 3.6 Gya.  The results are shown in Figure \ref{lifetime_fig} for 5 m/s surface winds, a run-off coefficient of 20\%, and a constant relative humidity of 50\%. Such a value for the humidity is, by definition, rather high for hyper-arid climates and therefore provides conservative (low) evaporation rates for the cold cases ($\overline{T_R}$<273K).  Similarly, Table \ref{table_precipitation_runoff_wind} shows the precipitation rates for a reference (arbitrary) atmosphere at T=273K, P=500 mbar, RH=50\% and different values for the surface wind speed and run-off coefficient.

\begin{table}
\scriptsize
\centering
\begin{tabular}{|c||c|c|c|}
\hline
Wind speed & 1 m/s &  5 m/s & 10 m/s \\
\hline
Evap. rate & 34 cm/yr  & 167 cm/yr & 334 cm/yr  \\
\hline
\hline
Prec. rate, $\alpha_{R}$=0\% &  35 cm/yr & 167 cm/yr &  334 cm/yr \\
\hline
Prec. rate, $\alpha_{R}$=20\% &  5 cm/yr & 24 cm/yr &  49 cm/yr \\
\hline
Prec. rate, $\alpha_{R}$=100\% &  1 cm/yr & 6 cm/yr &  11 cm/yr \\
\hline
\end{tabular}
\caption{Evaporation and precipitation rates as a function of the run-off coefficient and the surface wind speed. T=273K, P=500 mbar, RH=50\%}
\label{table_precipitation_runoff_wind}
\end{table}

Figure \ref{lifetime_fig} shows that for any pressure ranging between 7 mbar and 3 bar, an open water  lake (i.e an annually averaged regional surface temperature  $\overline{T_R}$>273K) will evaporate in less than 60 Earth years if not resupplied by some processes such as rain fall, snow melt, or groundwater flows. A direct consequence is that for these ranges of pressures and temperatures, a hydrological cycle is required to sustain open water lake conditions over a long period of time (>10,000 years). For cold climates ($\overline{T_R}\ll$273K) the lakes last longer without precipitation (in the order of hundreds to thousands of years), or alternatively maintain their levels constant with smaller precipitation rates (a few millimeters to a few centimeters per year).

When the temperature $\overline{T_R}$ is 273K and for a run-off coefficient of 20\%, Table \ref{table_precipitation_runoff_wind} shows that the precipitation rate estimates at Gale crater are in the order of a few tens of centimeter per year. These values lie toward the upper limit for the rainfall estimates required to form the valley networks during the Noachian (less than 1 mm yr$^{-1}$  to $\sim$ 1.1 m yr$^{-1}$  \citet{von_paris_2015}), and for reference, are similar to the range of those for the American south-western states \citep{NOAA_CCD_2015}. Ultimately this simple calculation only confirms the intuitive guess that in order to have Earth-like open water lakes on Mars, there is a requirement for high (Earth-like) rainfalls rates. Such high precipitation rates may be challenging to reconcile with the relatively low erosion at that time when compared to the Earth. Estimates for the erosion rates averaged over the entire Hesperian are 0.02-0.03 nm yr$^{-1}$ \citep{golombek_2006} and those for the Earth are $10^4-10^5$ nm yr$^{-1}$ \citep{carr_2010}. Another challenge with open water lakes in equatorial regions is that they may require large bodies of open water to power the hydrological cycle  (e.g. a lake in Eridania basin \citet{irwin_2002}  or a northern ocean  \citet{haberle_2017};  \citet{clifford_2001}; \citet{clifford_2017}). The implication would be that the temperatures must have been above the freezing point not only at equator, but also at higher latitudes so these large reservoirs would have also remained unfrozen.

In the following section, we test the  existence of open water lakes conditions by considering if first the annually-averaged \emph{global} mean temperature, second the annually-averaged \emph{equatorial} temperature  and third the equatorial \emph{seasonal} temperature could have been above  the freezing point.

\section{Temperature requirements for open water lakes}

\subsection*{Annually-averaged global mean temperature}

The simplest, yet most informative way to reproduce the Martian surface temperatures, or those of any other planetary body, is to compute the planetary equilibrium temperature $T_{eq}$ which represents the equilibrium temperature of a planet needed to balance the absorbed solar flux received from the sun in the absence of an atmosphere:

\begin{equation}
\underbrace{(1-a)\varphi_0 \pi R^2}_{received \; from \; sun}=\underbrace{\epsilon \sigma T_{eq}^4 4\pi R^2}_{emitted \; to \; space} \Rightarrow T_{eq}=\sqrt[4]{\frac{\varphi_0(1-a)}{4\; \epsilon \sigma}}
\label{eq_planet}
\end{equation}

with $a \approx 0.25$ the planetary albedo, $\varphi_0=\frac{1370 W.m^{-2}}{1.52^2} \approx 600 W.m^{-2}$ the solar flux at Mars, $\pi R^2$ the planet's cross section with R=3390 km the planet's radius, $\epsilon \approx 1$ the emissivity, $\sigma= 5.67 \times 10^{-8} W m^{-2}K^{-4}$ the Stefan constant and $4\pi R^2$ the surface area for the planet. The difference between the equilibrium temperature $T_{eq}$ and the effective (measured) surface temperature of the planet $T_{se}$ is attributed to greenhouse warming from the atmosphere. For the Earth, the greenhouse warming is about 33K, but for present-day Mars, it is on the order of a few kelvin only (see discussion in \citet{haberle_2013}). This may seem counter-intuitive, as the column amounts of CO$_2$ are $\approx 40$ times greater on Mars than they are on Earth. However the saturation of the main (15 $\mu$m) absorption band, the low atmospheric pressure (i.e weak pressure broadening) and the limited presence of other greenhouse gases, such as water vapor, make overall greenhouse warming less effective on Mars \citep{read_2015}. During the period of deposition at Gale crater, $\sim$3.6 Gya, stellar evolution models predict that the sun was about 75\% as bright as at present (faint young sun,  \citet{gough_1981};  \citet{kasting_1991}; \citet{bahcall_2001}).  Using this reduced value of incoming solar flux and the same value for the planetary albedo, the equilibrium temperature $T_{eq}$ from Eq.  (\ref{eq_planet}) drops to $\approx$ 196K implying that $\sim$ 77K of warming are needed to explain the presence of liquid water during the Hesperian ($\overline{T_G}$=273K).

\begin{figure}
\centering
\includegraphics[width=0.6\textwidth, trim=1.5cm 3 0cm 0 ,clip=true]{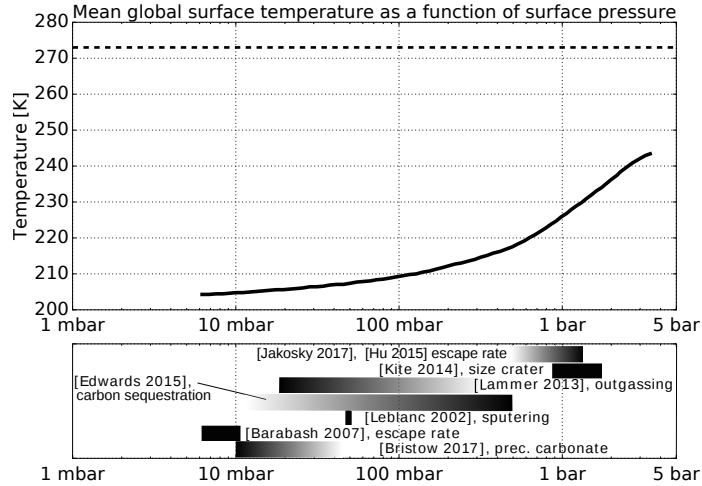} \\
\caption{(top) Surface mean temperature as a function of surface pressure for a CO$_2$-H$_2$O atmosphere, adapted from \cite{kasting_1991}'s 80 \% solar luminosity case. (bottom) Surface pressure estimates for early Mars derived from various methods.}
\label{kasting_fig}
\end{figure}

Recent estimates for ancient surface pressure on Mars range from a few tens of mbar \citep{bristow_2017},  possibly more if groundwater infiltration was occurring \citep{tosca_2018}, to approximately a bar \citep{barabash_2007,leblanc_2002,edwards_2015,kite_2014,jakosky_2017,hu_2015}.  However, limiting the early Mars paradigm to the CO$_2$/H$_2$O inventory alone is not sufficient, and \citet{kasting_1991} demonstrated  using a convective-radiative model that because higher pressures facilitate the condensation of CO$_2$, even 10 bars of CO$_2$ will not warm the atmosphere nearly close to the freezing point of water (see Figure \ref{kasting_fig}).  Many modeling studies have attempted to remedy \citet{kasting_1991}'s findings by including the effects of collision-induced absorption (CIA) and far-line absorption by CO$_2$ molecules \citep{wordsworth_2010}, using three dimensional simulations to model the radiative effects of CO$_2$ clouds \citep{forget_2013}, water ice clouds \citep{wordsworth_2013}, and dust \citep{kahre_2013}, simulating transient heating though impact craters \citep{segura_2012,steakley_2019}, testing the warming effects of sulfur dioxide (volcanism, \citet{halevy_2014}), H$_2$-driven greenhouse warming \citep{ramirez_2014,wordsworth_2017}, carbonate-silicate limit cycles \citep{batalha_2016}, and outgassing of methane \citep{kite_2017}. To date, finding ways to produce such an important warming remains one of the major challenges in planetary science, in the absence of an obvious mechanism for either transient or long-lived greenhouse atmospheres \citep{wordsworth_2016,haberle_2017}.

\subsection*{Annually-averaged equatorial temperature}

\begin{figure*}
\centering
\includegraphics[width=0.96\textwidth, trim=3cm 1cm 1cm 0cm,clip=true ]{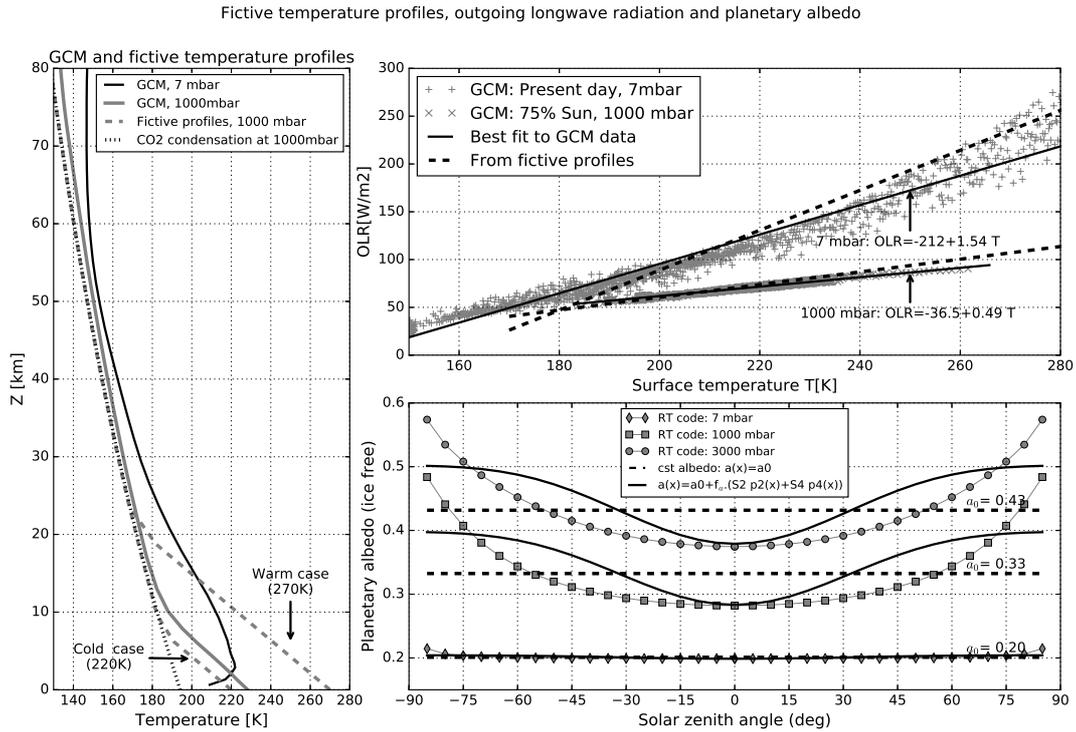} \\
\caption{(Left) Warm case and cold case fictive temperature profiles for a 1000 mbar pure CO$_2$ atmospheres (grey dashed lines). The CO$_2$ condensation curve for a 1000 mbar atmosphere (thin dotted line) shows how the profiles are constructed. Temperature profiles from full three-dimensional GCM simulations (year-average) are shown for reference for a present-day 7mbar atmosphere and a 1000 mbar ancient atmosphere (solid lines, see \citet{smith_2017} for others examples of temperature profiles derived from both modeling and observations). (Top right) Outgoing longwave radiation at the top of the atmosphere derived from the fictive temperature profiles and from the three-dimensional GCM simulations for the 7mbar and the 1000 mbar atmosphere. (Bottom right) Albedo as a function of the solar zenith angle (or equivalently as a function of the latitude for a 0$^o$ obliquity) for different surface pressures, are retrieved from the radiative transfer code (markers). The constant albedo model (dashed lines) and mean solar insolation-weighted model (solid lines) described in the Appendix are also shown.}
\label{OLR_IRD_fig}
\end{figure*}

In place of reducing the early Mars paradigm to a globally-averaged temperature requirement, we can assess whether the conditions for liquid water can be met locally.  Because the orbital parameters for Mars are known to have changed over time \citep{laskar_2002}, and since the atmospheric inventory during the Hesperian remains poorly constrained, we will use analytical formulations when possible so our calculations may be repeated for different climate scenarios.  The simplest way to estimate the regional temperature is to use a latitudinally-resolved energy balance model (EBM) to capture, at least to a first order, the dependence of the local energy balance on the latitude. Such models have been used for \emph{numerical} climate studies of Mars (e.g. \citet{postawko_1986}, \citet{hoffert_1981}, \citet{williams_1997}). In this paper though, we adapt one of the original methods by \citet{north_1975} to provide an \emph{analytical} solution for the annually-averaged surface temperature for early Mars for different surface pressure, eccentricity and obliquity scenarios. The reasoning for the energy balance model is similar to the equilibrium temperature calculation in Eq. (\ref{eq_planet}) where we said that globally, the absorbed solar radiation is balanced by the longwave flux emitted radiatively by the surface. The addition of an atmosphere to this framework has the following effects:  First, the outgoing longwave radiation (OLR) radiated to space is not equal the surface blackbody radiation anymore ($\sigma T_{sfc}^4$) but is now dependent on the radiative properties of the atmosphere (e.g. mass of the atmosphere, presence of greenhouse gases, clouds). Second, the atmosphere can transport diffusively the excess of heat (generally poleward) when the incident solar radiation and the OLR are unbalanced.  Let $E$ be the energy in an atmospheric column and assume that variations of $E$ can be related to variations of the surface temperature $T_s$. Following \citet{hoffert_1981}, we have:

\begin{equation}
\begin{split}
\frac{\partial E}{\partial t}= &\left(1-a(\Phi)\right)Q_{sun}(\Phi)-OLR(\Phi) \\
& +\dfrac{D}{cos(\Phi)}\dfrac{\partial}{\partial \Phi} \left(cos(\Phi)\dfrac{\partial T_{s}}{\partial \Phi} \right)
\end{split}
\label{ebm_eq}
\end{equation}
with $a$ the planetary albedo, $Q_{sun}$ the incident solar radiation, OLR the outgoing longwave radiation at the top of the atmosphere, $D$ the eddy diffusivity coefficient for the meridional transport and $T_{s}$ the surface temperature. Note that $a$, $Q_{sun}$ and the OLR are all dependent on the latitude $\Phi$. A convenient approximation is to assume that the OLR varies linearly with the surface temperature (more about this approximation below and in Figure \ref{OLR_IRD_fig}):

\begin{equation}
OLR= A +B \;T_{s}
\label{OLR_eq}
\end{equation}
with $A$ and $B$ two parameters to be determined for different atmospheric compositions and surface pressures. Setting the variable $x=sin(\Phi)$, averaging Eq. (\ref{ebm_eq}) over one Martian year (i.e. $\overline{\frac{\partial E}{\partial t}}\approx 0$) and re-organizing the terms lead to:

\begin{equation}
 \begin{split}
D\dfrac{d}{d x}\left((1-x^2)\dfrac{d \overline{T_s}(x)}{dx}\right)- &\left(A +B\;\overline{T_s}(x)\right) = \\
  & -\left(1-\overline{a}(x) \right) \overline{Q_{sun}}(x)
 \end{split}
\label{ebm_avg_eq}
\end{equation}

 which is a differential equation for the surface temperature $\overline{T_s}(x)$, the overbars denoting yearly-averaged values. In the particular case where the shortwave term $ \left(1-\overline{a}(x) \right) \overline{Q_{sun}}(x)$ in Eq. (\ref{ebm_avg_eq}) is approximated as a series of even Legendre polynomials, \citet{north_1975} showed that an analytical solution of Eq. (\ref{ebm_avg_eq}) exists and he provided a solution for the Earth with integration constants that depend on the boundary conditions at poles and  at the equator. We adapted the solution of  \citet{north_1975} to Mars in the special case where the albedo is a continuous function of the latitude, and we show in the Appendix that the annually-averaged temperature is:

 \begin{equation}
 \begin{split}
 \overline{T_s}(x)= &\frac{\varphi_0}{4 \sqrt{1-e^2}}\left(\frac{\mathit{Sa}_0}{B}\right)-\frac{A}{B} \\
  + & \frac{\varphi_0}{4 \sqrt{1-e^2}}\left(\frac{\mathit{Sa}_2 \; p_2(x)}{6D+B} +\frac{\mathit{Sa}_4 \; p_4(x)}{20D+B} \right)
 \end{split}
\label{ebm_sol_eq}
\end{equation}
with $x$ the sine of the latitude, $\varphi_0$ the solar constant at Mars (e.g. $\frac{1370}{1.52^2} \; W.m^{-2}$ for present day), $e$ the eccentricity, $D$ the diffusivity of the atmosphere, $A$ and $B$ the OLR parameters as in Eq. (\ref{OLR_eq}),  $\mathit{Sa}_n$ the insolation parameters provided in the Appendix and $p_n(x)$ the Legendre polynomials.  The diffusivity $D$ is estimated following \citet{hoffert_1981}:

\begin{equation}
D \sim 0.058 \left(\dfrac{p}{1013}\right) \Delta{\overline{T}_m} \quad [W m^{-2}K^{-1}]
\label{D_eq}
\end{equation}
with $p$ the surface pressure in unit of mbar and $\Delta\overline{T}_m$ the temperature difference between the poles and the equator. For present-day Mars ($p$=7mbar), $\Delta\overline{T}_m \approx 55K$ leads to $D\sim 0.02 W m^{-2}K^{-1}$. For an ancient, heavier, atmosphere we expect a stronger meridional transport ($D$ increases) because the atmosphere would re-distribute the heat more efficiently and therefore the temperature gradient $\Delta\overline{T}_m$ would decrease.  The former is unknown a priori but we can use a numerical solver to find values for $D$ and $\Delta\overline{T}_m = \overline{T_s}(1) -\overline{T_s}(0)$  that satisfy both Eq.  (\ref{ebm_avg_eq}) and Eq. (\ref{D_eq}). We found that the diffusivity $D$ ranges from 0.08 to 1.36 $W m^{-2}K^{-1}$ for 50 mbar- 4 bar atmospheres with present-day obliquity, and values for $D$ are given in Table \ref{table_OLR_IRD} in the Appendix  as a function of the surface pressure. We emphasize that since in our simple model, just about everything of what is currently known of the Martian atmosphere's dynamics is reduced to a single diffusivity parameter $D$, Eq. (\ref{D_eq}) is an order-of-magnitude estimate only. For a 1 bar atmosphere, this method gives D=0.70 $W/m^2/K$, which is 1.2-1.8 times the estimated value for the Earth at 1 bar ($D=$0.38-0.58$W m^{-2}K^{-1} $ \citet{north_1975b}; \citet{williams_1997}) and consistent with the scaling method proposed by \citet{williams_1997}.

Integrating Eq. (\ref{ebm_sol_eq}) latitudinally  across one hemisphere ($x$ ranging 0 to 1) and evaluating the function near the equator ($x \approx$ 0) provides estimates for the annually-averaged global mean temperature $\overline{T_{G}}$ and the annually-averaged regional temperature near Gale's latitude $\overline{T_{R}}$:

\begin{equation}
\begin{cases}
\begin{array}{l}
\displaystyle \overline{T_{G}}=  \frac{1}{B}\left(\frac{\varphi_0}{4 \sqrt{1-e^2}}\mathit{Sa}_0 -A\right) \\
\displaystyle \begin{aligned}
 \overline{T_{R}}= -\frac{A}{B}+ \frac{\varphi_0}{4 \sqrt{1-e^2}}\Bigg(\frac{\mathit{Sa}_0}{B}-& \frac{1}{2}\frac{\mathit{Sa}_2}{6D+B} \\
 +& \dfrac{3}{8}\frac{\mathit{Sa}_4}{20D+B} \Bigg)
 \end{aligned}
\end{array}
\end{cases}
\label{eq_TGTR}
\end{equation}

At this point, the missing parts of our climatic puzzle are the planetary albedo $\overline{a}(x)$, and the longwave parameters $A$ and $B$ that are used to estimate the outgoing longwave radiation. The OLR is strongly dependent on the surface pressure, the thermal structure of the atmosphere, the presence of condensable species (H$_2$O or CO$_2$ clouds), and the vertical distribution of radiatively active species (e.g. dust, water vapor, H$_2$, CH$_4$), which are all poorly constrained for early Mars. The effects of water ice clouds and CO$_2$ clouds on the longwave radiative budget (potentially positive feedback on the surface temperature) and on the planetary albedo (negative feedback on the surface temperature as more solar energy is reflected) are not included.  Under these assumptions, we found that adding small amounts of water vapor (a few tens to hundreds of precipitable $\mu$m)  in contrast to the much bigger water inventories used in other studies, (e.g 13.3 m global equivalent layer of water in \citet{segura_2012}) had little influence on the radiation budget for the range of surface temperatures considered ($\sim$200K-273K) thus we assume a pure CO$_2$ atmosphere.  Methods to estimate the radiative fluxes range in complexity from analytical single-band gray radiative models (as in \cite{robinson_2012}), to computationally expensive line-by line radiative codes (as in \citet{wordsworth_2017}). Some parameterization for the atmosphere's dynamical response to the radiative forcing may be included, for instance through a convective adjustment or by explicitly resolving advection and mixing, as it is the case in Global Climate Models (GCMs). For our purposes (linear fit of the OLR as a function of the surface temperature: $OLR= A +B \;T_{s}$), a rough estimate for the OLR  is sufficient. For simplicity, we construct a range of heuristic, annually-averaged, temperature profiles for different surface pressures and annual mean surface temperatures by simply following the dry adiabatic lapse rate ($\sim$ 4.7K/km) and ensuring that the temperatures stay above the CO$_2$ condensation curve at all heights.  Fictive temperature profiles for a cold case ($\overline{T_{R}}$=220K) and a warm case ($\overline{T_{R}}$=270K) are illustrated as dashed lines in Figure \ref{OLR_IRD_fig} (left)  and compared with the yearly-averaged prediction from a three-dimensional, fully interactive, GCM simulation with a 1 bar atmosphere and reduced solar luminosity (solid grey line). The fictive temperature profiles present the same general structure as the one predicted by the GCM, with an Earth-like troposphere and temperature gradually decreasing with height due to the increased efficiency of convective heat transport (mixing) at high pressures. This situation differs from present-day Mars (yearly-averaged values for a present-day, 7mbar GCM simulation are shown as a solid black line in Figure \ref{OLR_IRD_fig} (left)) where the vertical temperature structure varies widely as a function of the season and the local time \citep{savijarvi_1999}. The outgoing longwave radiation flux at the top of the atmosphere $Q^{TOA}_{IR\uparrow}$ and the downward infrared flux at the surface $Q^{sfc}_{IR \downarrow}$ are obtained by running the NASA Ames radiation code for each of these static  temperature profiles, including the effects of the infrared collision-induced absorption (CIA) and the far-line absorption at high pressure following \cite{wordsworth_2010}.  Figure \ref{OLR_IRD_fig} (top right) compares OLR estimates retrieved from the fictive profile method with OLR values from three-dimensional, fully interactive, GCM simulations for the 7mbar (present-day) atmosphere and the 1 bar atmosphere with reduced solar luminosity. The OLR values from the GCM simulations  (markers) are more dispersed than what the fictive profiles can account for since, for identical values for the surface temperature, air temperature inversions are not captured by the static profiles. However, the slopes for the OLR retrieved from the static profiles (dashed lines) match reasonably well the best fits to the GCM's outputs (solid lines). While being overly simplistic (the temperature structure is only an educated guess), this approach does capture the effect of increasing surface pressure and increasing atmospheric temperature on the longwave radiative budget.  Similarly, global mean values for the planetary albedo $a_0$ are retrieved from the radiative transfer calculation as a function of the surface pressure and are averaged over all the solar zenith angles, assuming a surface albedo of 0.2.  As it is instructive to account for the dependence of the albedo on the latitude, we provide an optional fitting factor $f_a$ that correlates high albedo values to high solar zenith angles (i.e low mean annual solar insolation),  does not alter the global mean value (the integral of $\overline{a}(x)$ over all latitudes is equal to $a_0$), and can be easily incorporated in the shortwave term of Eq. (\ref{ebm_avg_eq}) (see  bottom right panel in Figure \ref{OLR_IRD_fig}, and the discussion in the Appendix).

\begin{figure*}
\centering
\includegraphics[width=0.96\textwidth, trim=0cm 0 0cm 0 ]{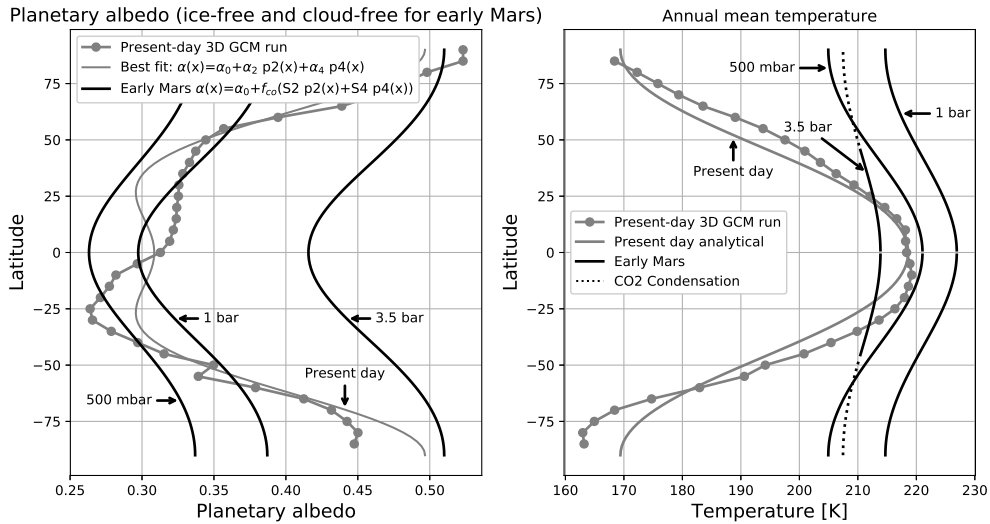} \\
\caption{(Left) Zonally-averaged planetary albedo from the 7mbar GCM simulation (grey markers), best fit for present-day Mars (grey symmetrical line with respect to the equator) and albedo distribution used for the early Mars predictions (black lines). (Right) Comparison of the mean annual temperature predicted by the NASA-Ames GCM (grey markers) with the analytical model for present-day Mars (grey line), and analytical predictions for early Mars for different surface pressures (black lines).  }
\label{EBM_fig}
\end{figure*}

We now have all the necessary tools to reproduce the first order, yearly-averaged climatology on Mars for different surface pressures. First, results from the analytical climate model are compared with those from the NASA Ames GCM for present-day Mars. The Nasa Ames GCM integrates a two-stream radiative transfer code, explicit 3D modeling of the atmospheric circulation, and a much more accurate representation of the aerosols and interactions between the surface of the atmosphere (see \citet{haberle_2019} for model updates). OLR values are retrieved from a 7 mbar GCM simulation (atmospheric dust and clouds are radiatively active), and fitted as a linear function of the surface temperature: $OLR= -212 +1.54\;T_s$ (see Figure \ref{OLR_IRD_fig}, top right). The diffusivity is $D\sim 0.02 W m^{-2}K^{-1}$, and we provide a direct fit to the NASA Ames General Circulation model for the annual mean, zonally-averaged planetary albedo:

\begin{equation}
\begin{cases}
\begin{array}{l}
\displaystyle \overline{a}(x)=a_0+a_2\;p_2(x)+a_4\;p_4(x)\\
\displaystyle a_0=0.329 \quad a_2= 0.0954 \quad a_4=0.0721\\
\end{array}
\end{cases}
\label{best_fit_7mbar_alb_eq}
\end{equation}
with $x$ the sine of the latitude, $p_n(x)$ the even Legendre polynomials. Note that the asymmetrical "W-shape" of the planetary albedo distribution retrieved from the GCM (grey markers in  Figure \ref{EBM_fig} (left)) can be related to Mars' surface albedo (refer to albedo maps by \citet{christensen_2001}), the annual deposition of CO$_2$ ice at high latitudes, and the distribution of clouds predicted by the GCM.  All together, these processes  lead to a yearly-averaged global mean value for the planetary albedo of $\sim$ 0.3, which is higher than the albedo obtained with the fictive profile method ($a_0$= 0.2 at 7mbar in Figure \ref{OLR_IRD_fig} (bottom right)). This means that the albedo values used in the latitudinally-resolved energy balance model  are favorable (low) to produce warm surface temperatures.  Figure \ref{EBM_fig} (right) shows that, despite its simplicity, the analytical climate model (grey line) reproduces remarkably well the annually-averaged surface temperature from the GCM for present-day Mars (grey markers and line). The temperatures predicted are almost identical to the GCM at the equator, typically 5-7K colder than the GCM at mid latitudes and slightly warmer than the GCM at the poles. Since only even Legendre polynomials are used to describe the distributions for the albedo and the solar insolation, the temperatures predicted by the analytical model are symmetrical with respect to the equator and can be seen as the average value of  both hemispheres.

The analytical model was then used to investigate different surface pressure, obliquity and eccentricity scenarios for early Mars. As an example, for a 1 bar, early Mars case, the luminosity $\varphi_0$ is reduced by 75\%, we read in Table \ref{table_OLR_IRD}: $D\sim 0.70 W m^{-2}K^{-1}$, $A$=-72, $B$=0.66, $\alpha_0 = 1- 0.33$ and $f_{co}=+0.135$ (see Appendix). Seasonal variations in temperatures are not considered in the annually-averaged EBM so the stability of the atmosphere against atmospheric collapse is not ensured but we can at least test the annual mean surface temperatures against the condensation temperature of CO$_2$ to check for the formation of perennial ice caps. The expression from \citet{forget_2013}; \citet{fanale_1982} for the condensation temperature of CO$_2$ is repeated here for convenience:
\begin{equation}
T_{cond}= \dfrac{-3167.8}{ln(0.01*P)-23.23}\quad [K]
\label{co2_condense_eq}
\end{equation} with $P$ the pressure in Pa and $P \lesssim$ 5 bars.

Figure \ref{EBM_fig} (left) shows the planetary albedo (ice-free and cloud free Mars) and annually-averaged surface temperatures (right) for a 500mbar, 1 bar and 3.5 bar atmospheres using present-day orbital parameters and 75 \% reduced solar luminosity (black lines).  For the 500 mbar atmosphere and for the 1 bar atmosphere, greenhouse warming and rather low values for the planetary albedo lead to surface temperatures warmer than present-day, despite a faint young sun (black lines in Figure \ref{EBM_fig} (right)). However, further increasing the surface pressure to 3.5 bar ultimately results in lower surface temperatures. This is due to atmospheric scattering at high pressure that increases the planetary albedo (left panel in Figure \ref{EBM_fig}) beyond what greenhouse warming can compensate for. Furthermore, surface temperatures beyond 50 degrees of the equator for the 3.5 bar case drop bellow the condensation temperature of CO$_2$ (dotted lines in Figure \ref{EBM_fig} (right)) so this solution is not stable against atmospheric collapse, consistently with the upper limit for stable atmospheres of $\sim$ 3 bar described in \citet{forget_2013}.
Using lower values for the obliquity and higher values for the eccentricity (not shown) raises the annually-averaged equatorial temperature by a few kelvin (up to $\sim$ 235K for a 2 bar atmosphere) but we found that there is no obvious combinations of orbital parameters that would lead to annually-averaged equatorial temperature $\overline{T_{R}}$ above 273K. There are two main reasons for this:
First, increasing the eccentricity $e$ increases the energy received by the whole planet (note the $\sqrt{1-e^2}$ factor in the denominator of Eq. (\ref{eq_TGTR}) and see discussion in \cite{berger_1994}), but this effect is minor: the planet experiences warmer temperatures near perihelion, colder temperatures near aphelion and the integrated effect over an entire year is small. Second, we notice that the meridional gradient in temperature  decreases significantly with increasing pressures (compare the pole-to-equator temperature differences between the 7 mbar case and the 1 bar case in Figure \ref{EBM_fig} (right)). This is again due to the increased efficiency in diffusive heat transport,  that better redistributes heat across all latitudes for high surface pressures  ($D$ increases in Eq.  (\ref{eq_TGTR})). As a result, although changes in obliquity do impact the latitudinal distribution in temperature, they do so to a much lesser extent than would be the case for Amazonian  (low surface pressure) climates. Therefore, we use present-day values for the orbital parameters for the rest of this study. The global effects of clouds, water vapor, or other reduced greenhouse gases on the longwave radiative budget are not included in this study but could easily easily incorporated into the parameters $A$ and $B$ \textbf{\footnote{the radiative transfer solver used in this study is freely available on the NASA Ames Mars Climate Modeling Center's website}}. Admittedly, several approximations in this model could be improved (e.g. better parameterization for the planetary albedo as a function of the latitude, use of self-consistent vertical temperature profiles, address the seasonal formation of ice caps), especially if the energy balance in Eq. (\ref{ebm_eq}) is solved numerically. Yet,  the analytical model is consistent with other GCM studies in predicting cold climates: For example \citet{wordsworth_2013} and \citet{forget_2013} found that the annual mean temperatures for equatorial regions peak from $\overline{T_{R}} \sim$230K ($\overline{T_{G}} \sim$ 210K-215K ) for 0.2-0.5 bar simulations to $\overline{T_{R}} \sim$245K ($\overline{T_{G}}\sim$230K) for 1-2 bar simulations.  This suggests that the surface of the lakes at Gale crater during the Hesperian must have been frozen, at least for part of the year.

\subsection*{Seasonal maximum for the equatorial temperature}

To assess if seasonal melting of the frozen lake occurred, we compute the seasonal variations in temperature that equatorial regions may have experienced.  Following \citet{dundas_2010}; \citet{hecht_2002}; \citet{kite_2013} ; \citet{mansfield_2018} , we use a computationally inexpensive one-dimensional numerical model with a single, parameterized, atmospheric layer and write the energy balance at the surface as:

\begin{equation}
\frac{\partial U}{\partial t}= (1-a) Q^{TOA}_{sun \downarrow}+Q_{IR \downarrow} +Q_{cond}+Q_{geo} +\mathit{SH} - \epsilon \sigma T^4_{surf}
\label{energy_budget_eq}
\end{equation}
with $U$ the energy per unit of area for the surface layer, $a$ albedo combining the reflection from the atmosphere and the ground (we use the constant value $a_0$ from Table \ref{table_OLR_IRD} in the Appendix as the average over all zenith angles), $Q^{TOA}_{sun \downarrow}$ the shortwave radiation at the top of the atmosphere and equal to 75\% of its present-day value, $Q^{sfc}_{IR \downarrow}$ the downward longwave radiation from the atmosphere, $Q_{cond}$ the conductive flux through the surface, $Q_{geo}$ the geothermal flux,  $\epsilon \sigma T^4_{surf}$ the longwave upward radiation, and $\mathit{SH}$ the sensible heat lost to the atmosphere. All the terms from Eq. (\ref{energy_budget_eq}) are described in the Appendix, as well as the physical parameters (e.g wind speed) chosen in the calculation. Note that the energy balance in Eq. (\ref{energy_budget_eq}) is applied to the surface layer this time, and not to the entire atmospheric column as in Eq. (\ref{ebm_eq}). With the visible atmospheric optical properties incorporated into the albedo $a$, almost of all the terms in Eq. (\ref{energy_budget_eq}) are either related to the surface layer itself (e.g. conduction to the ground, radiative losses) or to the atmospheric layer immediately adjacent to the surface (e.g. sensible heat flux). The only exception is $Q^{sfc}_{IR \downarrow}$ that depends on the temperature structure through the entire atmospheric column. We use a constant value $Q^{sfc}_{IR \downarrow}(\overline{T_{R}})$ with $\overline{T_R}$ derived from the analytical model (Eq. (\ref{eq_TGTR})) and we provide a parameterization for $Q^{sfc}_{IR \downarrow}$ in Table \ref{table_OLR_IRD} as a function of the surface pressure. As we are interested in the regional average values, the soil is assumed to be dry (no water/frost) so the ground properties are representative of the Martian regolith and there is no latent heat terms. The time-stepping model in Eq. (\ref{energy_budget_eq}) was tested against predictions from the NASA Ames GCM for present-day Mars and reproduced reasonably well the surface temperatures from the Ames GCM: over a full Martian year, the annually-averaged surface temperature at Gale crater $\overline{T_{R}}$ was 223K both models and the daily-averaged temperature from the one dimensional model was no more than 3K different than the GCM's at any time of the year. The one dimensional model also reproduced adequately the phase and amplitude of the diurnal ground temperature variation (peak afternoon temperatures and nighttime temperatures were within 5K of those predicted by the GCM), in agreement with ground temperature measurements from the \textit{Curiosity} rover (see \citet{martinez_2017} for a overview of ground temperature data as a function of the local time).

Then, we ran the time-marching model with different values for the surface pressure and a reduced solar luminosity.  Consistently with the predictions from the analytical climate model, we verified that changes in obliquity and eccentricity only resulted in warmer/colder departures from an annual mean value that was tens of kelvin below 273K. Therefore we keep present-day orbital parameters for simplicity and focus instead on the effects of the surface pressure on the amplitudes of the diurnal and seasonal variations in surface temperature.

\begin{figure}
\centering
\includegraphics[width=0.5\textwidth, trim=0cm 0 0cm 0 ]{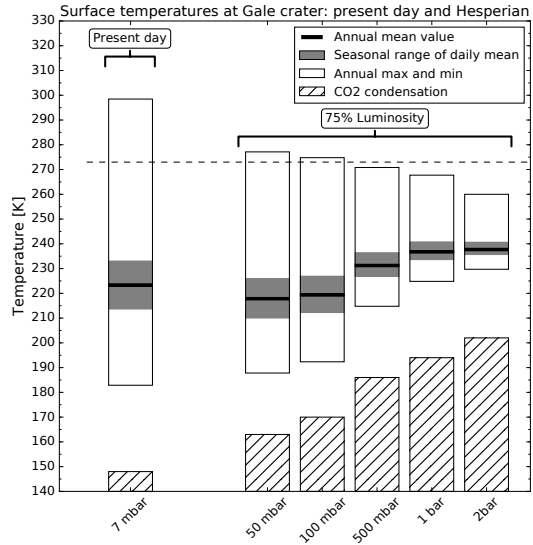} \\
\caption{Estimates for the annually-averaged regional temperature near Gale's latitude (thick black lines), seasonal variation of the daily-averaged temperature (grey bars), yearly temperature maximum and minimum (white bars) and condensation temperature of CO$_2$ (hatched bars) from the time-stepping model for different surface pressures.}
\label{sfc_ebm_1D_fig}
\end{figure}

Figure \ref{sfc_ebm_1D_fig}  confirms that annual mean surface temperatures $\overline{T_{R}}$ (black lines) increase with increasing pressure as the atmosphere provides more greenhouse warming, up to a certain point where the increase in albedo (atmospheric scattering) limits this effect ($\overline{T_{R}}$ does not increase much from 1 bar to 2 bar). The time-stepping calculation shows that the seasonal maximum temperature (e.g average temperature for the hottest day of the year, shown as the upper parts of the grey bars in Figure \ref{sfc_ebm_1D_fig}) are well below the freezing point (dashed line at 273K). This means that not only the baseline state (year-average) for the lakes was frozen, but furthermore the lakes must have been \emph{perennially} ice-covered. However, not unlike present-day Mars, peak daytime temperatures (e.g temperature for the hottest hour of the year, shown as the upper parts of the white bars in Figure \ref{sfc_ebm_1D_fig}) do get above 273K for the moderate pressure cases (<500mbar). Note that increasing the surface pressure tends to decrease the daily and the seasonal ranges of temperature (respectively white and grey bars in Figure \ref{sfc_ebm_1D_fig}), toward a less extreme, Earth-like configuration where convective eddies efficiently cool the radiatively-heated surface during the day (see Eq. (\ref{eq_SH}) in the Appendix and \citet{rafkin_2013}). The general trend is that low to moderate surface pressures (P<500 mbar) result in cold mean temperatures (T $\ll$ 273K) but large seasonal and daily excursions from the mean, and high surface pressures (P>500 mbar) result in slightly higher mean values but smaller daily and seasonal excursions from the mean.

To summarize, we consistently demonstrated that there is no combination of surface pressure and orbital parameters that could have led to either permanent or seasonal open water lakes at Gale crater during the Hesperian. Our best estimates for the equatorial temperature falls short of at least 30K to the freezing point, using pure ice as a limiting case (absent of salt).  Since climate studies have consistently predicted a cold climate for early Mars \citep{haberle_2017,wordsworth_2016}, we propose to investigate the cold and wet scenario where lakes created during the Noachian could have persisted in the form of perennially ice covered-lakes.

\section{The ice-covered lakes hypothesis}

\subsection*{Analogy with perennially ice-covered lakes in Antarctica}

In the Dry Valleys of Antarctica, the annual mean temperatures are below freezing, yet sustained liquid water occurs under perennial ice covers \citep{armitage_1962,doran_1994,sokratova_2011}. These perennially ice-covered lakes occur in topographic lows and are either isolated or abutted by glaciers. The seasonal melting of snow and/or glacial meltwater recharges the liquid water column of the lakes and provides a source of energy in the form of latent heat. The thickness of the ice cover is set by the energy balance and by the mass balance of the ice cover, ice being removed by ablation at the surface and added to the bottom by freezing \citep{mckay_1985}. Two different types of perennially ice-covered lakes are possible analogues for the lake that could have been present in ancient Gale crater: 1) lakes that are recharged entirely by inflowing melt streams, such as the lakes in the McMurdo Dry Valleys and 2) lakes that have significant recharge from subaqueous melting of abutting glaciers, such as Lake Untersee \citep{steel_2015,faucher_2019}.  Sedimentary structures have been documented at the bottom of both types of lakes \citep{simmons_1985,squyres_1991,levitan_2011}. For the McMurdo Dry Valleys lakes, maximum air temperatures above freezing (expressed as degree-day above freezing) are required in summer for meltwater to flow into the lake. Applying this analogue to Mars, \citet{mckay_1991} explored the idea that sustained liquid water could have been persisted long after the global mean temperature fell below freezing. In their study, the requirement for long-lasting lakes was that the peak seasonal temperatures exceeded the freezing point of water. Under these conditions, they found that sustained liquid water on Mars could have been preserved for 100's of millions of years after the global mean temperature fell below the freezing point.

Lake Untersee, East Antarctica, has a surface area of 11.4 km$^2$. It is sealed by a $\sim$ 3m perennial ice cover and dammed by the Anuchin Glacier \citep{andersen_2011,wand_1997,steel_2015,faucher_2019}. The annual mean air temperature is -10.6C$^o$ \citep{andersen_2015} The bottom of Lake Untersee is dominated by clay-sized sediment (although larger grains have been documented) and the mud in the lake is primarily derived from the abutting glacier \citep{steel_2015,faucher_2019}. At Lake Untersee, solar radiation penetrates the ice cover and melts the glacier's wall underneath the lake's surface, recharging the lake in meltwater. As a follow-up study of \citet{mckay_1991} and \cite{fairen_2014} in which glacial and periglacial activity is present, we propose to investigate the Lake Untersee model for Gale crater because it does not requires liquid water to flow at the surface at any time of the year. While this model alone may not explain all of the sedimentary deposits encountered by the \textit{Curiosity} rover (e.g. river deposition), it provides a starting point for exploring the range of possible lake scenarios for Gale crater during the Hesperian that are consistent with climatic studies.

Solar radiation and the ice albedo are key parameters allowing for the preservation of ice-covered lakes in below-freezing temperatures \citep{mckay_1985,mckay_2004}.  Since the values for the albedo given in Table \ref{table_OLR_IRD} already include reflection by the ground, the reduction in solar insolation due to atmospheric reflection is conveniently defined through a mean visible optical depth $\tau$ in order to differentiate the effects of the atmosphere from the effect of the local ice on the solar energy available at the surface:
\begin{equation}
Q^{sfc}_{sun \downarrow}=Q^{TOA}_{sun \downarrow} e^{-\tau}
\label{eq_Qsun_tau}
\end{equation}

\begin{figure}
\centering
\includegraphics[width=0.5\textwidth, trim=0cm 0 0cm 0 ]{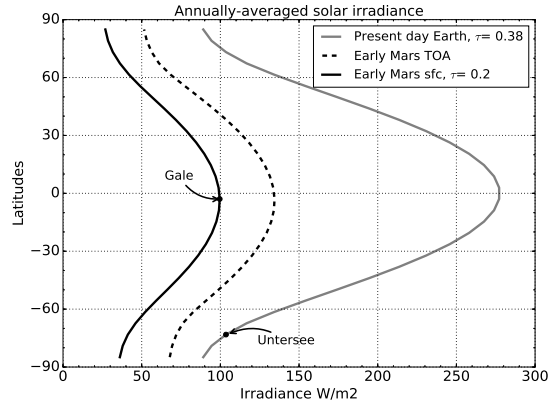} \\
\caption{Annually-averaged surface solar irradiance as a function of latitude for early Mars and present-day Earth under clear sky conditions. The irradiance at the top of the atmosphere (TOA) and at the surface for a visible optical depth $\tau$=0.2 are shown for Mars and $\tau = 0.38$ is used for Earth. (See \citet{liu_1960} for a simple parameterization for the diffuse radiation for the Earth) }
\label{solar_irradiance_fig}
\end{figure}

Following \citet{levine_1977}; \citet{hottel_1976}; \citet{liu_1960}, we show in Figure \ref{solar_irradiance_fig} the annually-averaged clear sky solar irradiance at the surface as a function of latitude for early Mars with present-day orbital parameters, and for present-day Earth (we use a visible optical depth $\tau=0.2$ for a 500 mbar atmosphere). The  annually-averaged solar irradiance at Gale crater and at Lake Untersee are comparable:  the faint young sun and the greater sun-Mars distance result in a reduction of the solar irradiance  with respect to present-day Earth by a factor of 3 (we have $0.75\; { \left(\frac{1}{1.52}\right)}^2   \approx \nicefrac{1}{3} $), but the near-equatorial latitude of Gale crater results in an enhancement of the solar energy at the surface, which offsets that reduction. It is therefore plausible that, at least for certain configurations of obliquity and eccentricity,  perennially ice-covered lakes similar to the ones observed at high-latitudes in Antarctica persisted in equatorial regions at Mars during the Hesperian.

\subsection*{Ice thickness calculation for Hesperian lakes}

We compute the energy budget for a perennially ice-covered lake located inside Gale crater. For the sake of simplicity, we assume a regional forcing of the atmospheric temperatures above the lake, meaning that the temperature at the surface of  the lake is known and prescribed by the energy budget over the large, mostly dry area, that surrounds Gale crater (i.e. Eq. (\ref{energy_budget_eq}) and Figure \ref{sfc_ebm_1D_fig}). Since the ice cover sits on top of a liquid water column, the temperature at the bottom of the ice cover is also known and equal to 273K. Thus, with the top and bottom temperatures of the ice cover fully constrained, the energy balance can be used to derive the thickness of the ice cover. The ice thickness equation is adapted from \citet{mckay_1985}:

\begin{equation}
\overline{Z} =\frac{b \;ln(\frac{T_o}{\overline{T_R}}) +c \left(\overline{T_R}-T_o\right)-F_{sun} h \left(1-e^{\frac{-\overline{Z}}{h}}\right)}{F_{geo}+\rho_{ice} v L(1-\gamma_{\text{glacier}})}
\label{eq_ice_thick1}
\end{equation}

with $\overline{Z}$ the annual mean value for the ice-cover thickness, $b$, $c$ defining the ice thermal conductivity (see Table \ref{ice_table} in the Appendix), $T_o$=273K the temperature at the ice/liquid interface, $\overline{T_R}$ the annually-averaged regional temperature, $F_{sun}$ the effective solar energy available at the top of the ice cover, $h$ the attenuation length of solar radiation within the ice, $\rho_{ice}$ the ice density, $v$ the freezing rate at the bottom of the ice cover and equal to the sublimation rate in Eq.  (\ref{eq_combinedRH}), $L$ the latent heat of fusion for water, $F_{geo}$ the geothermal flux and $\gamma_{\text{glacier}}$ a new term that represents the fraction of water input to the lake provided as melted ice from the glacier.

When $\gamma_{\text{glacier}}$ is set to zero, the lake is entirely resupplied by summer melt or subglacial flows, Eq. (\ref{eq_ice_thick1}) is the same as in \cite{mckay_1985}, and the term $ \rho_{ice} v L$ corresponds to the energy released at the bottom of the ice cover as the water freezes. When $\gamma_{\text{glacier}}$ equals one, no latent heat term appears in Eq. (\ref{eq_ice_thick1}): latent heat is still released at the bottom of the ice cover when the water freezes, but an equal amount of energy was needed to first melt that same volume of ice from the glacier so the net contribution is zero. When $\gamma_{\text{glacier}}$ is different from 0 or 1, we must verify that the mass balance for the ice cover is still valid. Let $v_f$ be the freezing rate at the bottom of the ice cover, $v_{\text{liquid}}$, the deposition rate for water brought into the lake by summer melt or subglacial flows and $v_{\text{melted}}$ the deposition rate for water brought into the lake by the ice melted from the glacier. Then, the total mass of water frozen at the bottom of the ice cover over one year is:

\begin{equation}
\begin{split}
\underbrace{\rho_{ice} v_f A}_{\text{frozen bottom}} = & \underbrace{\rho_{ice} v_{\text{liquid}} A}_{\text{provided by melt streams}} +\underbrace{\rho_{ice} v_{\text{melted}} A}_{\text{provided by glacier}} \quad [kg/yr]\\
& = \rho_{ice}(1-\gamma_{\text{glacier}})vA + \rho_{ice}\gamma_{\text{glacier}}vA \\
&= \rho_{ice} v A
 \end{split}
 \label{eq_mass_balance_in_out}
\end{equation}

We have $v_f=v$ so ablation from the surface is still balanced by freezing at the bottom of the ice cover and mass is conserved. The energy released per unit area at the bottom of the ice cover when water freezes is $\rho_{ice} v_f L$ =$\rho_{ice} v L$,  the energy used to melt the glacier per unit area is $\rho_{ice} v_{\text{melted}} L= \rho\gamma_{\text{glacier}}v L$ and both terms appear in the denominator of Eq. (\ref{eq_ice_thick1}). Following \citet{mckay_2004}, we use a simplified two-band model to define the optical properties for the ice: albedo, transmissivity, are assumed to be uniform in the visible and ice is considered opaque to wavelengths larger than 700 nm.  This approximation is justified by both measurements and theory \citep{mckay_1994,mckay_2004,bohren_1983}, and the effective solar energy available to melt the ice is:

\begin{equation}
F_{sun}=f_{700}(1-a_{ice})Q^{sfc}_{sun \downarrow}
\label{eq_Qsun_sfc}
\end{equation}
 with  $f_{700}$ the fraction of solar energy for wavelengths below 700nm (photosynthetically active radiation or PAR) that penetrates the ice cover, $a_{ice}$ the visible albedo over the ice, and $Q^{sfc}_{sun \downarrow}$ the solar flux at the surface from Eq. (\ref{eq_Qsun_tau}) and Figure \ref{solar_irradiance_fig}. Note that since ice is mostly opaque to infrared radiation, we expect the ice cover thickness to decrease for atmospheres with strong greenhouse warming because of the higher ambient temperatures (i.e. lower conductive flux driving the freezing of the ice and higher ablation rates) and not directly because of the increased atmospheric infrared back radiation as the latter is not deposited at the bottom of the ice cover.

We show in the Appendix that the solution for the annual mean thickness $\overline{Z}$ in Eq. (\ref{eq_ice_thick1}) is:

\begin{equation}
\begin{cases}
\begin{aligned}
 &\overline{Z}= h\left(W\left(\frac{e^{\frac{\beta}{\alpha}}}{\alpha}\right) -\frac{\beta}{\alpha}\right)\\
 &\alpha=\frac{F_{geo} +\rho_{ice} v L \left(1-\gamma_{\text{glacier}}\right)}{F_{sun}}\\
 &\beta= 1- \frac{b \;ln(\frac{T_o}{\overline{T_R}}) +c \left(\overline{T_R}-T_o\right)}{F_{sun} h}
\end{aligned}
\end{cases}
\label{eq_analytic_mckay85_k_var}
\end{equation}

with $W$ the Lambert-W function. Using Eq. (\ref{eq_analytic_mckay85_k_var}), the annual mean thickness $\overline{Z}$ is computed for different combinations of ice extinction path lengths and mean annual surface temperatures. The mean equatorial surface temperatures $\overline{T_R}$  range between ~230K (weak greenhouse warming due to CO$_2$ only, see Figure \ref{sfc_ebm_1D_fig}) to 270K (up to +40K of warming due to some unspecified radiatively active species, e.g. a reducing H$_2$ or CH$_4$ atmosphere as in \cite{wordsworth_2017}). It is interesting to consider the specific case of clear ice in Antarctica as the ice has to be relatively transparent for this model to work. The 3 m thick ice cover on Lake Untersee has a visible albedo of 0.66 and a PAR transmissivity $T_{PAR}=(1-a_{ice} )e^{\nicefrac{-z}{h}}$ of 0.05 (\cite{andersen_2011}), implying an extinction path length $h$ = 1.6 m. On Lake Hoare, the average value of $h$ is $\sim$ 1 m for the entire ice cover, but for the lower layers of clear ice $h$ = 2.3 m (an extinction coefficient of 0.435 m$^{-1}$)  since attenuation near the surface of this ice cover is due primarily to accumulation of airborne dust. We consider extinction path lengths ranging from $h$=0.5m to $h$=6.25 m for very clear ice (an extinction coefficient of 0.16 m$^{-1}$, see \citet{mckay_1994}).

\begin{figure*}
\centering
\includegraphics[width=0.9\textwidth,clip=true, trim=1.5cm 0cm 0cm 0 ]{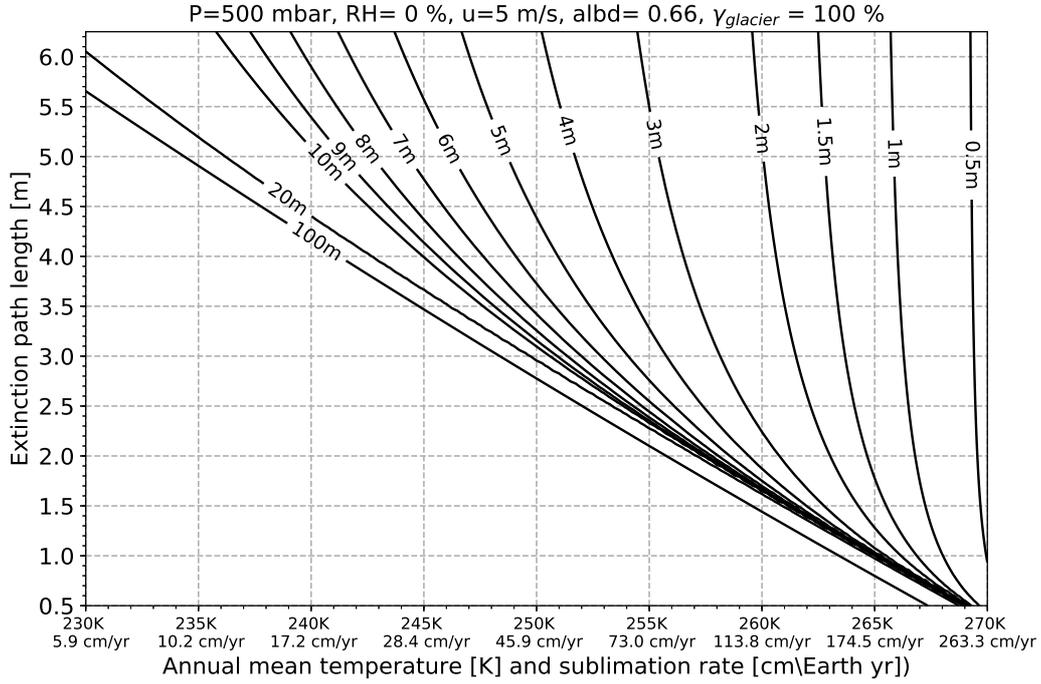} \\ 
\caption{Ice thickness as a function of the extinction path length and mean annual surface temperature for a glacier-fed lake}
\label{ice_thickness_2D_fig}
\end{figure*}
The thickness of the ice cover is shown in Figure \ref{ice_thickness_2D_fig} for a 500 mbar atmosphere, assuming surface winds of 5 m/s, an albedo of 0.66, a dry atmosphere (the relative humidity is $RH$=0\%), and for the limiting case where the lake is entirely resupplied by subaqueous melting of the glacier ($\gamma_{\text{glacier}}$=1). Figure \ref{ice_thickness_2D_fig}  shows that for a mean annual temperature of 230K (weak greenhouse warming due to CO$_2$ only), there is no thin ice solution, that we define as the cases where the ice thickness is less than 10 meters. This is comparable to the range of values for the perennially ice-covered lakes in Antarctica, typically 3-6 m \citep{mckay_1985,doran_1994,wand_1997}. However, when the ice is very clear ($h$=5m) thin ice solutions exist for annual mean temperatures as low as 240K, even though the maximum annual temperature never reaches 273K. At 255K, the requirement on the extinction path length for a 10 m thickness decreases to $h$=2.4m and an ice cover thickness similar to Lake Untersee (3 m) is obtained for very clear ice ($h$=5.5m). To summarize, extinction path lengths >2.4m are required in order to have a lake system resupplied entirely by subaqueous melting of a glacier for these very cold temperatures (<240K-255K). This requires the ice to be generally clearer than the ices for the lakes in Antarctica and also results in thicker ice covers (e.g. 10 m vs $\sim$ 3-6 m for the Antarctic lakes).
For clean ice covers such as Lake Untersee and the lower ice layers of Lake Hoare the main source of opacity is scattering due to the presence of bubbles trapped in the ice \citep{mckay_1994}. These bubbles are formed due to the supersaturation of the lake water in gas (mostly nitrogen and oxygen \citet{priscu_1999}; \citet{andersen_2011}). Therefore, the ice transparency for Hesperian perennially-covered lakes would have been intrinsically correlated with the saturation level for the lake's water, and extinction path lengths  $h \gg$ 2.3 m are plausible if the bubble content for the ice was low.  Figure \ref{ice_thickness_2D_fig} is a limiting case because we assumed that the lake was entirely resupplied by subaqueous melting of the glacier. The existence of subglacial flows would allow to partially recharge the lake without ever contacting the atmosphere (no requirement for T>273K) while inducing no penalty on the energy balance in Eq. (\ref{eq_ice_thick1}) since the water comes as liquid and does not require prior melting of the ice. Furthermore, we saw in Figure \ref{sfc_ebm_1D_fig} that for even for moderate surface pressures ($P$<500 mbar) the daily maximum temperature can reach the freezing point even though the mean annual temperatures stay as cold as 220K. Therefore, it is reasonable to assume that even if the seasonal temperatures stayed below freezing, a fraction of the lake water could have been resupplied by daytime summer-melt, as it is the case in the Dry Valleys in Antarctica \citep{mckay_1985}.

We will now consider intermediate scenarios where the lake is resupplied by both subaqueous melting of the glacier, subglacial flows and summer meltwater. For a given mean ice thickness, the fraction of water supplied by melting from the glacier $\gamma_{\text{glacier}}$ in Eq. (\ref{eq_ice_thick1}) is easily derived as a function of the ice cover properties (extinction path length, albedo), and the climatology (solar insolation, regional temperature $\overline{T_R}$, ablation rates $v$):

\begin{equation}
\begin{split}
\gamma_{\text{glacier}}=1&+\frac{F_{geo}}{\rho_{ice} v L} \\
&-\frac{b \;ln(\frac{T_o}{\overline{T_R}}) +c \left(\overline{T_R}-T_o\right)-F_{sun} h \left(1-e^{\frac{-\overline{Z}}{h}}\right)}{\rho_{ice} v L \overline{Z}}
\end{split}
\label{eq_gamma_glacier}
\end{equation}

Eq. (\ref{eq_gamma_glacier}) shows the maximum amount of lake water that can be resupplied by subaqueous melting of the glacier given a set of extinction path length and surface temperature, and for two thicknesses of interest, $\overline{Z}$=3m and $\overline{Z}$=10m.

\begin{figure}
\centering
\includegraphics[width=0.9\textwidth,clip=true, trim=1.5cm 0cm 0cm 0 ]{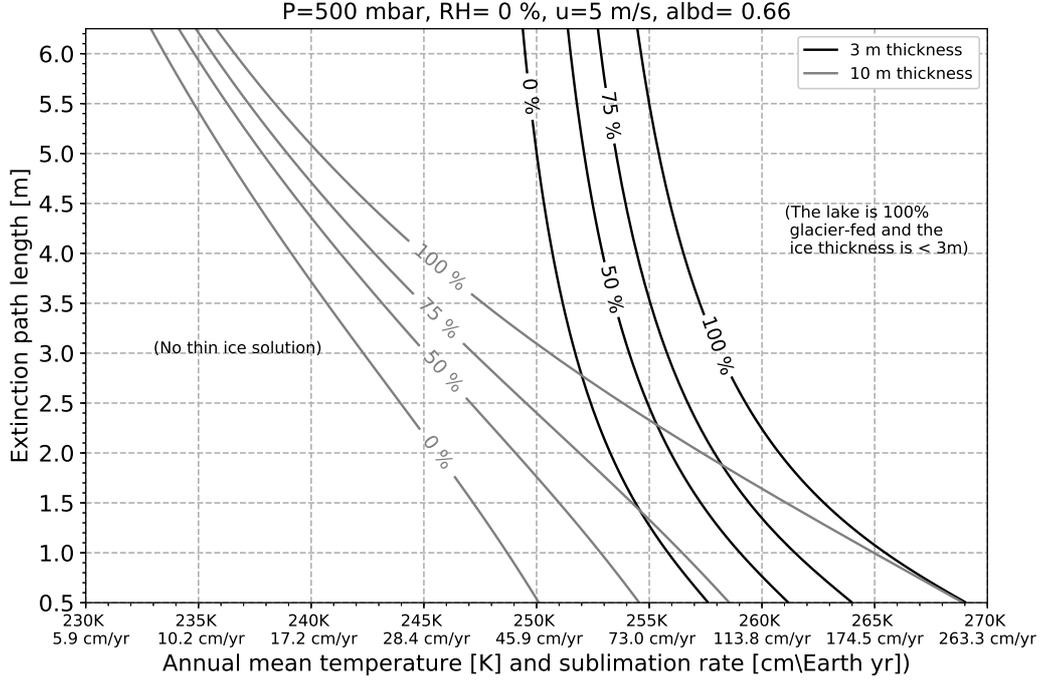} \\
\caption{Percentage of the lake's water that can be resupplied by a glacier while keeping the ice cover thickness under 3 meters (black) and  10 meters (grey)}
\label{glacier_thickness_fig}
\end{figure}

When only a $\nicefrac{1}{4}$ of the lake water's is resupplied by summer melt or subglacial flows and for a conservative value for the extinction path length $h$=2m, the temperature requirement for thin (<10m) ice covers drops from $\overline{T_{R}}\sim$ 258K to $\overline{T_{R}}\sim$ 252K (grey $75\%$ contour in Figure \ref{glacier_thickness_fig}). Similarly, for that same extinction path length $h$=2m, ice covers as thin as 3 meters are possible at 253K  when all the water is provided as summer melt (black $0\%$ contour in Figure \ref{glacier_thickness_fig}).

Providing that the ice is sufficiently clear (extinction path length $h$>2 m), the ice-covered lake model can explain how a liquid environment could have been sustained at Gale crater through the Hesperian. In fact, the perennially ice-covered lakes could have subsisted as long as a source of meltwater (here a glacier) was available to resupply the lakes without the need for mean annual temperatures above freezing which no climate model has been able to reproduce. As a rough estimate, for the equatorial temperatures  $\sim$ 10-15K above the global mean value (see Figure \ref{EBM_fig}), mean annual temperatures at Gale crater $\overline{T_{R}}$=240-255K correspond to global mean temperatures $\overline{T_{G}} \sim$ 230-245K.  The latest results from \citet{wordsworth_2017} suggest that this range of temperature is readily obtained with 0.5 - 1 bar of CO$_2$ and $\sim$ 1-5\% H$_2$, with less or no reducing species at all if the surface pressure is greater than 1 bar.

\subsection*{Proposed geological context and implications for the carbonate content of the sediment}
\begin{figure}
\centering
\includegraphics[width=0.8\textwidth ]{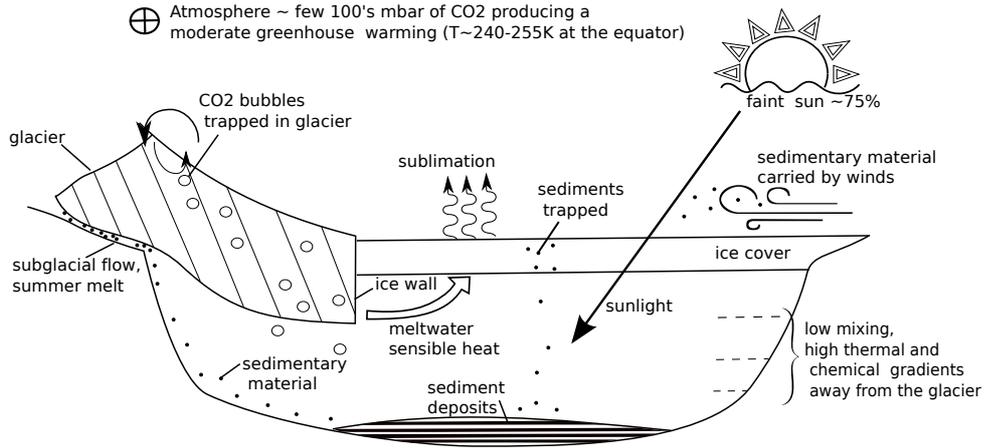} \\
\caption{A cold hydrological system at Gale crater during the Hesperian}
\label{lake_sediments_fig}
\end{figure}

Based on these results, we propose the geological context illustrated in Figure \ref{lake_sediments_fig} for Hesperian lakes with: a) an ice cover that limits the evaporation losses and seals the lake from the atmosphere b) a glacial wall, subglacial flows or episodic melt streams which resupply the lake with meltwater, c) sediment material being transported through the ice cover from the base of the glacier.

The MSL mission discovered that the carbonate content of the sediment is low, typically below the detection limit of the CheMin instrument of $\sim$ 1 wt\% \citep{ming_2014}. The nondetection of carbonate in the sediment in Gale crater is an unexpected outcome in the context of sediment formed in open water lakes that are in thermodynamic equilibrium with a CO$_2$ atmosphere, since sustaining unfrozen lakes requires a  substantial amount of CO$_2$ to provide enough greenhouse warming (\citet{bristow_2017}  and note that groundwater infiltration was proposed as a plausible mechanism to produce magnetite and H$_2$ in high $p_{CO_2}$ atmospheres \citet{tosca_2018})	On Earth, perennially ice-covered lakes are not necessarily in equilibrium with the atmosphere. For instance Lake Untersee is supersatured in nitrogen, oxygen, and,  interestingly, deficient in CO$_2$ \citep{wand_1997}. This is due to its perennial ice cover which isolates (seals) the lake from the environment, and to the presence of microbial communities that consume the CO$_2$ dissolved in the water \citep{townsend_2008,andersen_2011}. In our model the sublimation of the ice cover is balanced by melting of the glacier, so any CO$_2$ trapped within the glacier will eventually be dissolved in the water.

The carbon flux entering the lake and available for carbonate precipitation is therefore still related to the atmospheric pressure, however it is now controlled by both the sublimation of the ice cover and the CO$_2$ bubble content of the glacier. With carbon itself making about 27\% of the mass of a CO$_2$ molecule, the carbon flux going into the lake is can be estimated as:

\begin{equation}
F_{C}= 0.27 \; F_{ice} \; V_{bubble} \; \rho_{CO2} \quad [kg_{C}/m^2/yr]
\end{equation}

with $F_{ice}$ in $[kg_{ice}/m^2/yr]$ the mass flux of ice from the glacier melting into the lake and distributed over the surface area of the lake,  $(V_{bubble} \; \rho_{CO2})$ the mass of CO$_2$ trapped in one kg of ice with $V_{bubble}$ the bubble content of the glacier in $[m^3/kg_{ice}]$ and $\rho_{CO2}$ in $[kg_{CO2}/m_{ice}^3]$ the density of CO$_2$ bubbles. Furthermore, the density  $\rho_{CO2}$ is related to the atmospheric pressure $P_{CO2}$:

\begin{equation}
 \rho_{CO2} = \frac{P_{CO2} M_{CO2}}{R T}
\end{equation}
with $M_{CO2}$= 0.044 $[kg/mol]$ the molecular mass of CO$_2$, $R$=8.314 $[J/kg/mol]$ the gas constant and T the air temperature. Note that in the case where the lake is entirely resupplied by subaqueous melting of a glacier, the ice mass flux from the glacier $F_{ice}$ exactly balances the sublimation rate of the ice cover $E_{combined}$ from Eq. (\ref{eq_combinedRH}), written in units of $[kg_{ice}/m^2/yr]$. For instance, with $F_{ice}$=600 $[kg/m^2/yr]$ (ablation rate of 60 cm per Earth year, see Figure \ref{lifetime_fig}), a relatively low value for the surface pressure $P_{CO2}$ =50 mbar, T=255K, and a typical bubble content for Antarctic glaciers $V_{bubble}=10^{-4}[m^3/kg_{ice}]$ \citep{craig_1993}, we found a carbon flux entering the lake equal to  $F_C \sim 1.6g/yr/m^2$.  We adapted \citet{bristow_2017} calculations to account for the reduced input of carbon into the water with respect to open water lakes and  estimate a new  calcite content for the sediment. Assuming that all the carbon ends up as calcite in the sediment so that the area of sediment matches the area of ice cover, a sediment density of 3 $g/cm^3$, a porosity of 0.6, a carbon flux of $1.6g/yr/m^2$, and a sedimentation rate through the ice cover of 0.1 cm yr$^{-1}$, we found that the calcite  makes $\sim$ 1 wt\% of the sediment, which is consistent with the detection limit of CheMin. Since the carbon flux $F_C$ scales linearly with the atmospheric pressure and the bubble content, higher surface pressures require bubble contents lower than those typically found in the Antarctic glaciers in order to stay below the 1 wt.\% limit (e.g  $V_{bubble} \leqslant 10^{-5}[m^3/kg_{ice}]$ for $P_{CO2}$ =500 mbar).  Furthermore, if the CO$_2$ carried into the lake in these glacier bubbles is largely converted to carbonate then there would be less buildup of dissolved gases in the water column. This would reduce the bubble formation in the ice cover and imply clear ice, which is favorable to the formation of a thin ice cover.

Perennially ice-covered lakes also allow for stratification in water temperature and chemistry as ice covers prevent wind-driven turbulence within the lakes \citep{obryk_2016,townsend_2008,priscu_1999}. Lake Untersee is damned by a glacier so cooling of the water near the glacial wall induces a buoyancy-driven circulation that mixes the lake where the glacier reaches \citep{steel_2015,faucher_2019}. However, part of the lake is shielded from these currents and remains stratified \citep{bevington_2018}. It is worthwhile to reconsider the nature of the lake deposits observed in Gale crater in the context of an ice-covered lake. We hypothesize that the intrinsic physical stability of a perennially ice-covered lake would be generally consistent with the chemical stratification observed within the sediment in Gale crater \citep{hurowitz_2017}.

Finally, we would like to comment on the potential of perennially ice-covered lakes as habitats for biology. In Antarctica, microbial communities sometimes produce thick mats at the lake bottom of perennially ice-covered lakes, including Lake Untersee  \citep{andersen_2011,hawes_2013,mackey_2015,sumner_2016}. Likewise, if microbial communities were present on early Mars, perennially ice-covered lakes might have provided a stable, long-lived shelter for their persistence and it is plausible that cosmic rays would have mostly erased their traces from the shallow Martian subsurface \citep{pavlov_2012,montgomery_2016}. Nevertheless, if perennially ice-covered lakes did exist on Mars, sites like Gale crater may be ideal places to search for biosignatures, a few meters beneath the surface.

\section{Summary and caveats for the ice-covered lake hypothesis}

The ice-covered lake model provides an interesting way to decouple the mineralogy and the climate by limiting the gas exchanges between the sediment and the CO$_2$  atmosphere, and it eliminates the requirement for annual mean temperatures above the freezing point. We foresee that moderate surface pressures (<1 bar of CO$_2$) would still result in a low calcite content within the sediment while allowing for enough greenhouse warming ($\overline{T_{R}} \sim$ 240K-255K) to prevent the lakes from freezing to the bedrock. In Gale crater, it is primarily the presence of well cemented and laterally extensive laminated mudstones that suggests a lacustrine environment \citep{grotzinger_2014,grotzinger_2015}. On Earth, laminated mud may be deposited from inflowing streams but also by abutting glaciers in perennially-covered lakes, as is the case for Lake Untersee \citep{andersen_2011,levitan_2011,rivera_hernandez_2019}. However, contrary to what had been observed in the sediment deposits at Gale crater, the laminated mud at Lake Untersee also contains disseminated medium sand to granule size sediment, with rare fine to medium gravel \citep{rivera_hernandez_2019}. At Lake Untersee, the sand at the bottom of the lake originates from the ice cover: the lake ice acts as a trap for aeolian sediment and the sand then penetrates the ice cover either by melting the ice or migrating downward through conduits such as ice-grain boundaries, ice cracks, or gas bubble channels (e.g. \citet{jepsen_2010} and references therein). If the sand on a perennially ice-covered lake at Gale crater did not migrate through the ice (either due to the absence of conduits in the ice cover or the impossibility of melting its way through the ice), this would result in the deposition of laminated mud from an abutting glacier without disseminated sand from the ice cover.

Admittedly, the Lake Untersee model may not explain all of the lake deposits in Gale crater, in particular, those  that are interpreted to have a fluvial origin and require running water \citep{williams_2013,grotzinger_2014,grotzinger_2015,anderson_2015,edgar_2017}. We saw in Figure \ref{sfc_ebm_1D_fig} that peak daytime temperatures reach 273K for pure CO$_2$ atmospheres with moderate surface pressures ($P$<500mbar) but this study has focused mostly on the steady-state climates and we did not address specifically the erosion potential of episodic snowmelt (see \citet{kite_2013}). Ephemeral streams and rivers form in the cold and hyperarid McMurdo Dry Valleys and induce erosion, deposition and the formation of alluvial fans and deltas \citep{head_2014,shawn_1980}. Notably, the seasonal flow patterns for rivers in the Dry Valleys show a strong diurnal cycle associated with warmer daytime temperatures \citep{shawn_1980}. This make them compelling analogues for Gale crater, as equatorial regions on Mars typically experience strong seasonal temperature cycles (due to obliquity and eccentricity effects), and large (tens of kelvin) diurnal temperature cycles.  Gale crater also contains a handful of geologic landforms that have been interpreted as possible evidence for past cold-climate environments such as decametric polygons \citep{oehler_2016}, and  lobate features \citep{fairen_2014}.

Conversely, the lack of unambiguous evidence for a cold climate at Gale crater has been pointed out: frost wedges, glacial tills, and glaciogenic sedimentary deposits (e.g. coarse cobbles, boulder conglomerates) are typically absent from MSL's observations \citep{grotzinger_2015}, which is a weakness for the Lake Untersee model. If a cold-based glacier was present in Gale crater, it is plausible that its presence in the rock record would be subtle if sliding did not occur at the rock-ice interface.  On Earth cold-based glaciers have low erosion potential (compared to temperate glaciers) and can sometimes preserve the landscape rather than erode bedrock \citep{marchant_2007,richardson_1996}. Similarly, one can  hypothesize that freeze-thaw weathering would be fairly limited in a cold environment that experiences only episodic melting rather than regular freezing/melting cycles. Rather than refuting the picture of open water lakes that has emerged from detailed sedimentological observations by the MSL science team, we support \citet{fairen_2014} who raised a fundamental question: how much of the stratigraphy observed at Gale crater could possibly be interpreted with a cold hydrological cycle?

\section{Appendix}


\subsection*{A. Derivation of the free evaporation}

The evaporation resulting from free convection is (\cite{ingersoll_1971}; \cite{adams_1990}; \cite{dundas_2010}):
\begin{equation}
E_{free}= \frac{1}{\rho_{ice}}K_M \frac{M_{H_20}}{R}\left(\frac{e^{surf}_{sat}}{T_{surf}} -RH \frac{e_{sat}^{atm}}{T_{atm}} \right) \quad [m/s]
\label{eq_freeRH}
\end{equation}

Introducing  $D=D_{H_2O/CO_2} [m^{2}.s^{-1}]$  the diffusion coefficient of water vapor into the CO$_2$ atmosphere and $\nu [m^2s^{-1}]$ the kinematic viscosity evaluated at $\frac{1}{2}(T_{surf}+T_{atm})$, \citet{adams_1990} defined the bulk coefficient $K_M$ as:
\begin{equation}
K_M=0.14{\left(\frac{{D }^2 B}{\nu}\right)}^{\frac{1}{3}} \quad [ms^{-1}]
\label{eq_KM}
\end{equation}
with $B$ the buoyancy force per unit mass resulting from the presence of H$_2$O in the denser CO$_2$ atmosphere:
\begin{equation}
B=g \frac{\rho_{atm}-\rho_{surf}}{\rho_{surf}}  \quad [ms^{-2}]
\label{eq_B}
\end{equation}

 with $\rho_{atm}$ and $\rho_{surf}$ the atmospheric densities of the CO$_2$ /H$_2$O mixture of the ambient air and at the surface, respectively. We have:

\begin{equation}
\begin{split}
\frac{\rho_{atm}-\rho_{surf}}{\rho_{surf}} & = \Bigg( \left(\frac{P^{atm}_{CO_2}M_{CO_2}}{RT_{atm}}+\frac{e^{atm} M_{H_20}}{RT_{atm}}\right)\\
& -\left( \frac{P^{surf}_{CO_2}M_{CO_2}}{RT_{surf}}+ \frac{e^{surf}_{sat} M_{H_2O}}{RT_{surf}}\right) \Bigg)\\
& \times \frac{1}{ \frac{P^{surf}_{CO_2}M_{CO_2}}{RT_{surf}}+ \frac{e^{surf}_{sat} M_{H_2O}}{RT_{surf}}}
\end{split}
\label{eq_Drho_rho1}
\end{equation}

with $P_{atm}$ the total pressure, and $P_{CO_2}$ the partial pressure of CO$_2$. Replacing $P_{CO_2}=P_{atm}-e$  respectively at the surface and in the atmosphere and rearranging (\ref{eq_Drho_rho1}) we found:

\begin{equation}
\begin{split}
\frac{\rho_{atm}-\rho_{surf}}{\rho_{surf}} & = \Bigg( M_{CO_2}P_{atm}\left(\frac{T_{surf}}{T_{atm}}-1 \right)\\
&- \frac{T_{surf}}{T_{atm}}(M_{CO_2}-M_{H_20})e^{atm} \\
& + (M_{CO_2}-M_{H_2O})e^{surf}_{sat} \Bigg)\\
& \times \frac{1}{M_{CO_2} P_{atm} -(M_{CO_2}-M_{H_2O})e^{surf}_{sat}}
\end{split}
\label{eq_Drho_rho2}
\end{equation}

Using $e^{atm}= RH \, e^{atm}_{sat}$, we obtain:
\begin{equation}
\begin{split}
\frac{\rho_{atm}-\rho_{surf}}{\rho_{surf}} & = \Bigg(M_{CO_2}P_{atm}\left(\frac{T_{surf}}{T_{atm}}-1 \right)\\
& +(M_{CO_2}-M_{H_2O})(e^{surf}_{sat}-\frac{T_{surf}}{T_{atm}}RH\,e^{atm}_{sat})\Bigg) \\
&\times \frac{1}{M_{CO_2} P_{atm} -(M_{CO_2}-M_{H_2O})e^{surf}_{sat}}
\end{split}
\label{eq_D_rho_rho3}
\end{equation}
Assuming $T_{atm} \sim T_{surf}$, we have:

\begin{equation}
\frac{\rho_{atm}-\rho_{surf}}{\rho_{surf}} = \frac{(M_{CO_2}-M_{H_2O})e^{surf}_{sat}(1-RH)}{M_{CO_2} P_{atm} -(M_{CO_2}-M_{H_2O})e^{surf}_{sat}}
\label{eq_D_rho_rho4}
\end{equation}

and we combine Eq. (\ref{eq_freeRH}), (\ref{eq_KM}) and (\ref{eq_D_rho_rho4}) and find:

\begin{equation}
\begin{split}
E_{free}=\frac{1}{\rho_{ice}} \underbrace{0.14 D{\left(\frac{g \frac{\rho_{atm}-\rho_{surf}}{\rho_{surf}}}{{\nu}^2}(\frac{\nu}{D})\right)}^{\frac{1}{3}}}_{K_M} \frac{M_{H_2O}}{R T_{surf}} e_{sat}(1-RH)\\
 [m.s^{-1}]
\end{split}
\label{eq_dundas_appendix}
\end{equation}

We used the formulations provided in \citet{altheide_2009} for $D_{H_2O/CO_2}(T,P)$, $e_{sat}(T)$,  $ \nu(T,P)$ to  obtain a formulation for the free evaporation rate as a function of the surface pressure, surface temperature and relative humidity. Note that Eq.  (\ref{eq_dundas_appendix}) is similar to the expression from \citet{altheide_2009} when $RH=0$, or \citet{ingersoll_1971} who assumed $\frac{\nu}{D} \sim 1$, $RH=0$, $\Delta_{\eta}=\frac{\rho_{sat}}{\rho_{ice}}$ and 0.17 close to 0.14.

Note that at high relative humidity, the evaporation rate decreases for two reasons. First the buoyancy $B$ decreases as the relative difference in density   $\frac{\rho_{atm}-\rho_{surf}}{\rho_{surf}}$ between the ambient air and the surface decreases (physically, this is due to the difference in molar masses between the denser CO$_2$ and the lighter H$_2$O since a mixture of  $CO_2+H_2O$ is lighter than CO$_2$ alone). Second,  the difference between the saturated vapor density in the ambient air and at the surface:
\begin{equation}
\frac{M_{H_20}}{R}\left(\frac{e^{surf}_{sat}}{T_{surf}} -RH \frac{e_{sat}^{atm}}{T_{atm}}\right)
\end{equation}
 decreases. Eventually, when the air is saturated (RH= 100 \%) the evaporation shuts off completely.

On the other hand, when the surface temperature is warmer than the ambient air $T_{atm} < sim T_{surf}$, an additional buoyancy force remains in Eq. (\ref{eq_D_rho_rho3}) and leads to higher evaporation rates.

\subsection*{B. Derivation of the wind-driven  evaporation}

We have \citep{adams_1990,dundas_2010}:
\begin{equation}
E_{forced}=\rho_{atm} \frac{\kappa u^{*}}{ln\left(\frac{z5}{z0}\right)}(q_{sat}-q_{atm})\quad [kg/s]
\label{eq_E_forced1}
\end{equation}
with $\rho_{atm}$ the atmospheric density, $\kappa  \sim 0.4 $ m s$^{-1}$ the Von Karman constant, $u^{*}$ a friction velocity and $q=\frac{M_{H_2O}}{M_{CO_2}}\frac{e}{P_{atm}}$ the specific humidity.
Assuming a log-wind profile for the boundary layer:
\begin{equation}
u=\frac{u^{*}}{\kappa}\left[ ln\left(\frac{z5}{z0}\right) +\phi \right] \quad[m/s]
\label{eq_ustar}
\end{equation}
with $\phi \sim 0$ for a neutral boundary layer, combining (\ref{eq_E_forced1}) and (\ref{eq_ustar}), we obtain:

\begin{equation}
E_{forced}=\frac{1}{\rho_{ice}} A \; \frac{M_{H20}}{R}\left(\frac{e^{surf}_{sat}}{T_{surf}} -RH\frac{e^{atm}_{sat}}{T_{atm}} \right) \quad[m/s]
\label{eq_forcedRH}
\end{equation}

and
\begin{equation}
A=\frac{{\kappa}^2}{ln\left(\frac{z5}{z0}\right)^2} \quad [m/s]
\label{eq_funct_forced}
\end{equation}

with $M_{H_2O}$ the molecular weight of water, $R$=8.314 [J/kg/mol] the gas constant, $\kappa=0.4$ the Von Karman's constant, $u$ the wind speed at a reference altitude above ground, $T_{atm}$ and $T_{surf}$ the air and surface temperatures, $e_{sat}$ and $e=RH e_{sat}$ the saturated and ambient vapor pressure of the air. We used the surface roughness $z0$=0.01m, and $z5$=5m as a reference altitude.

\subsection*{C. Derivation of the analytical solution for the energy balance model}

\citet{north_1975b} gives a particular solution to Eq. (\ref{ebm_avg_eq}), for the OLR :

\begin{equation}
OLR=\sum_{n, even}\dfrac{Q \;H_n(x_s) \; p_n(x)}{n(n+1)D'+1}
\label{north_1975_eq}
\end{equation}
where Q is a globally-averaged value for the solar flux ($Q=\dfrac{\varphi_0}{4}$ in our study), $x$ the sine of the latitude with $x_s$ the position of the discontinuity in albedo at the edge of an ice sheet (if any), $D'=\frac{D}{B}$ in our study, $H_n$ the insolation components and $p_n$ the Legendre polynomials. For Earth's present-day obliquity (and by analogy for Mars'present-day obliquity as $\beta_{E}=23^o \sim \beta_{M}= 25^o$), \cite{north_1975b} shows that expending Eq. (\ref{north_1975_eq}) to the first two terms is generally sufficient. In order to solve Eq. (\ref{ebm_avg_eq}) for a wide range of obliquities, we use the fourth degree approximation from \citet{nadeau_2017} for the mean annual insolation as a function of the latitude:

\begin{equation}
\begin{cases}
\begin{array}{l}
\displaystyle \overline{Q_{sun}}(x)= \frac{\varphi_0}{4 \sqrt{1-e^2}}\left(1+S_2 \;p_2(x)+S_4 \;p_4(x) \right)\\
\displaystyle S_2=-\tfrac{5}{8} p_2(cos(\beta))\\
\displaystyle S_4=-\tfrac{9}{64} p_4(cos(\beta))
 \end{array}
\end{cases}
\label{qsun_term_eq}
\end{equation}

with $\varphi_0$ the solar constant, $e$ the eccentricity, $\beta$ the obliquity and $p_n$ the Legendre polynomials:

\begin{equation}
\begin{cases}
\begin{array}{l}
\displaystyle p_0(x)=1\\
\displaystyle p_2(x)=\left(3x^2-1 \right)/2 \\
\displaystyle p_4(x)=\left(35x^4-30x^2+3 \right)/8 \\
\displaystyle p_6(x)=\left(231x^2-315x^4+105x^2-5 \right)/16 \\
\displaystyle p_8(x)=\left(6435x^8-12012x^6+6930x^4-1260x^2+35 \right)/128 \\
\end{array}
\end{cases}
\label{pol_legendre_eq}
\end{equation}

The planetary albedo, formally defined as the ratio of the reflected to upcoming shortwave radiations at the top of the atmosphere $a =\nicefrac{Q^{TOA}_{sun\uparrow}}{Q^{TOA}_{sun\downarrow}}$, can vary due to several factors: a) the reflectance of the soil (presence of CO$_2$ ice), b) the optical properties of the atmosphere (increased scattering at higher pressures, presence of CO$_2$ clouds), and c) the difference in the solar zenith angle at different latitudes (obliquity-dependent).

Therefore, it is desirable to use a formulation for the annually-averaged co-albedo $\overline{\alpha}(x)=\left(1-\overline{a}(x) \right)$ (i.e $\alpha_0 = 1-a_0$, $\alpha_2 =- a_2$, $\alpha_4 =- a_4$)  that will be appropriate for different climate scenarios. In place of the simple step function in \cite{north_1975} or the discontinuous second order approximation in \cite{north_1975b}, we use instead a continuous, fourth degree approximation for the co-albedo:

\begin{equation}
\overline{\alpha}(x) = \alpha_0 +\alpha_2\;p_2(x)+\alpha_4\;p_4(x)
\label{co_albedo_eq}
\end{equation}
where the $\alpha_n$ coefficients can be, by order of increasing complexity for the parameterization:
\begin{itemize}
\item[--] reduced to a global mean value $\alpha_0$ (i.e. $\alpha_2=\alpha_4=0$). $\alpha_0$ is retrieved from the radiative transfer calculations for different surface pressures (the albedo increases with increasing pressure due to scattering) and averaged over all zenith angles
\item[--] approximated as a combination of a pressure-dependent term $\alpha_0$ and the distribution of the incident shortwave radiation which is related to the annually-averaged value for the solar zenith angle:
\begin{equation}
\overline{\alpha}(x)=\alpha_0+ f_{co} \left(S_2 \;p_2(x)+S_4 \;p_4(x) \right)
\label{co_albedo_f_eq}
\end{equation}  with $\alpha_2=f_{co}\times \;S_2$, $\alpha_4=f_{co} \;S_4$, and $f_{co}$ a fitting factor for the co-albedo  with  $f_{co}=-f_{a}$ and $f_{a}$ the optional albedo factor from Table \ref{table_OLR_IRD}
\item[--] directy fitted to observations or modeling results from a Mars Global Climate Model (MGCM) where the polar caps are resolved and the dependence of the albedo on the solar zenith angle is explicitly computed.
\end{itemize}

Using equations (\ref{qsun_term_eq}-\ref{co_albedo_eq}), we can show that:

\begin{equation}
\begin{cases}
\begin{array}{l}
\displaystyle \overline{\alpha}(x) \overline{Q_{sun}}(x)= \frac{\varphi_0}{4 \sqrt{1-e^2}}\sum_{n=0, even}^{n=8} \mathit{Sa}_n\; p_n(x)\\
\displaystyle \mathit{S_a}_0=\tfrac{1}{5}S_2\alpha_2+\tfrac{1}{9}S_4\alpha_4 +\alpha_0\\
\displaystyle \mathit{S_a}_2=S_2\alpha_0+\tfrac{2}{7}S_2\alpha_2+\tfrac{2}{7}S_2a4+\tfrac{2}{7}S_4\alpha_2+\tfrac{100}{693}S_4\alpha_4+\alpha_2\\
\displaystyle \mathit{S_a}_4=\tfrac{18}{35}S_2\alpha_2+\tfrac{20}{77}S_2\alpha_4+S_4\alpha_0+\tfrac{20}{77}S_4\alpha_2+\tfrac{162}{1001}S_4\alpha_4+\alpha_4\\
\displaystyle \mathit{S_a}_6=\tfrac{5}{11}S_2\alpha_4+\tfrac{5}{11}S_4\alpha_2+\tfrac{20}{99}S_4\alpha_4\\
\displaystyle \mathit{S_a}_8=\tfrac{490}{1287}S_4\alpha_4 \\
\end{array}
\end{cases}
\label{Qsun_co_a_eq}
\end{equation}

where $S_n$ are the solar terms, $\alpha_n$ are the co-albedo terms, and $p_n$ the Legendre polynomials. The higher order terms can be neglected (${S_a}_8={S_a}_6 \sim 0$), and we obtained an expression for the annually-averaged solar radiation absorbed by an atmospheric column:
\begin{equation}
\overline{\alpha}(x) \overline{Q_{sun}}(x)= \frac{\varphi_0}{4 \sqrt{1-e^2}}\left(\mathit{S_a}_0+\mathit{S_a}_2 p_2(x)+\mathit{S_a}_4 p_4(x)\right)
\label{Qsun_co_a_simple_eq}
\end{equation}
with $\varphi_0$, $\beta$, $e$, $\mathit{S_a}_0$, $\mathit{S_a}_2$, $\mathit{S_a}_4$  dependent on the climate scenarios considered for early Mars (e.g. high obliquity, high eccentricity, presence of ice, reduced solar luminosity $\varphi_0$).

The formulation in Eq. (\ref{Qsun_co_a_simple_eq}) for the absorbed shortwave radiation is a continueous function of the latitude, which is a special case in \cite{north_1975b} (ice-free case, $x_s$=1). It follows that $H_n(1)=\mathit{Sa}_n$, and using $T=\frac{OLR-A}{B}$ in Eq. (\ref{north_1975_eq}) we can finally write the solution of the energy balance differential equation as:

 \begin{equation}
 \overline{T_s}(x)= \frac{\varphi_0}{4 \sqrt{1-e^2}}\sum_{n=0, even}^{4}\dfrac{\mathit{Sa}_n \; p_n(x)}{n(n+1)D+B}-\dfrac{A}{B}
\label{ebm_sol_appendix_eq}
\end{equation}

\begin{table*}
  \centering
  \small
  \begin{tabular}{|c|c|c|c|c|c|c|c|}
    \hline
    \multirow{2}{*}{Pressure [mbar]}& \multirow{2}{*}{D [$W/m^2/K$]}& \multicolumn{2}{c|}{$Q^{TOA}_{IR\uparrow}(T)= A+B\;T$} & \multicolumn{2}{c|}{$Q^{sfc}_{IR\downarrow}= A_1+B_1\;T$}& \multicolumn{2}{c|}{Albedo} \\
    \cline{3-8} & & A & B & $A_1$ & $B_1$ & $a_0$ & $f_a$ \\
    \hline
    7 & 0.02&-212&1.54 &-176 &0.96 &\multicolumn{2}{c|}{(see text)}\\ \hline
   50& 0.08&-274&1.78&-255&1.42&0.21&-0.033\\ \hline
  100& 0.14&-248&1.63&-286&1.61&0.23&-0.051\\ \hline
  200& 0.25&-208&1.40&-300&1.75&0.25&-0.075\\ \hline
  500& 0.46&-134&0.99&-341&2.04&0.29&-0.111\\ \hline
  800& 0.61&-93&0.77&-350&2.13&0.32&-0.128\\ \hline
 1000& 0.70&-72&0.66&-355&2.17&0.33&-0.135\\ \hline
 1500& 0.86&-40&0.49&-351&2.20&0.37&-0.143\\ \hline
 2000& 1.00&-19&0.39&-335&2.15&0.39&-0.146\\ \hline
 2500& 1.11&-6&0.32&-327&2.13&0.41&-0.146\\ \hline
 3000& 1.20&5&0.27&-318&2.10&0.43&-0.144\\ \hline
 3500& 1.28&13&0.23&-310&2.08&0.45&-0.142\\ \hline
 4000& 1.36&19&0.20&-303&2.06&0.46&-0.139\\ \hline
 5000& 1.49&30&0.15&-289&2.02&0.49&-0.134\\ \hline
  \end{tabular}
  \caption{Estimates for the outgoing longwave radiation at the top of the atmosphere, downward infrared flux at the surface, and planetary  albedo as a function of the surface pressure derived from the \textit{fictive} temperature profiles (clear sky, no clouds, see Figure \ref{OLR_IRD_fig}). The coefficients of determination $R^2$ for the linear fits to the OLR $Q^{TOA}_{IR\uparrow}$ and to the downward infrared flux at the surface $Q^{sfc}_{IR\downarrow}$  are greater than 0.95 and 0.88, respectively. For the planetary albedo (total reflection by atmosphere and the surface), we assume a surface albedo of 0.2. The global mean value for the co-albedo $\alpha_0$ and fitting factor $f_{co}$ are easily related to the global mean albedo $a_0$ and the optional albedo factor $f_{a}$ using $\alpha_0= 1-a_0$ and $f_{co}=-f_{a}$. Estimates for the diffusivity $D$ from Eq. (\ref{ebm_sol_eq}) are also given in the table.}
  \label{table_OLR_IRD}
\end{table*}

\subsection*{D. Terms used in the time-dependent surface energy balance model}

\subsubsection*{Downward shortwave radiation}

The downward shortwave radiation $Q^{TOA}_{sun \downarrow}$ is derived from \citet{levine_1977,allison_2000}.

\begin{equation}
\begin{cases}
\begin{array}{l}
\displaystyle Q^{TOA}_{sun \downarrow}=Q^{mean}_{sun} C_{Ls} C_{lat} \\
\displaystyle Q^{mean}_{sun}=0.75\frac{1370}{1.52^2}\\
\displaystyle C_{Ls}= \left(\dfrac{1+e \, cos(L_s-L^P_s)}{1-e^2}\right)^2\\
\displaystyle C_{lat}=\left(sin(\phi) \, sin(\delta)+cos(\phi)cos(\delta)cos(\frac{2 \pi t_{noon}}{T_{Mars}}) \right)
\end{array}
\end{cases}
\label{eq_sun}
\end{equation}
 $Q^{mean}_{sun}$ is the mean solar flux at Mars $\sim$ 3.6 Gya fixed to 75\% of today's value with 1370 $Wm^{-2}$ the solar constant and $1.52$ the distance to the Sun in astronomical units, $ C_{Ls} $ is a correction factor for the variations of the Sun-Mars distance with the solar longitude $L_s$, $e$ is the eccentricity,  $L^p_s=248$ the solar longitude at perihelion.  $C_{lat} $ accounts for the solar zenith angle with  $\phi$ the latitude, $\delta$ the solar declination with $sin(\delta)=sin(\beta)sin(L_s)$ with $\beta$ Mars' obliquity, $t_{noon}$ the time elapsed after solar noon and $T_{Mars}=88775s$ the length of a Martian solar day. See \citet{levine_1977} for the integration of the incident solar irradiance over a day including atmospheric attenuation.

\subsubsection*{Downward longwave radiation at the surface}

A rough estimate for the annually-averaged downward infrared radiation at the surface is also obtained from the radiative transfer calculations and available in Table \ref{table_OLR_IRD}. As for the OLR, we use a simple linear fit to the radiative transfer output:
\begin{equation}
\overline{Q^{sfc}}_{IR \downarrow} = A_1 + B_1 \; \overline{T_R}
\end{equation}
with $\overline{T_R}$ the annually-averaged regional temperature obtained from Eq. (\ref{eq_TGTR}).
\subsubsection*{Geothermal flux}
The geothermal flux $Q_{geo}$ is used as an energy input at the base of the model. We assumed a value $Q_{geo}$=30 mW/m$^{-2}$ following \citet{dundas_2010}.  Note that this energy input is small relatively to the downward shortwave and longwave fluxes ($Q_{geo}\ll Q_{IR \downarrow}$, $Q_{sun}$).

\subsubsection*{Atmospheric temperature and sensible heat}

For the atmospheric temperature, we use the parametrization from \citet{dundas_2010}:
\begin{equation}
T_{atm}=T_{min}^{b}T_{surf}^{1-b}
\end{equation}

with $T_{min}$ the most recent diurnal minimum of the surface temperature, $T_{surf}$ the current surface temperature and $b$ an atmospheric parameter.

For the sensible heat, we use the formulation from \citet{dundas_2010}; \citet{williams_2008}:

\begin{equation}
\begin{cases}
\begin{array}{l}
\displaystyle SH\Bigr|_{\substack{free}}=0.14(T_{atm}-T_{sfc})k_{atm}{\left( \frac{C_P\, \rho_{atm}}{k_{atm}}\frac{g \frac{\rho_{atm}-\rho_{surf}}{\rho_{surf}}}{{\nu}}\right)}^{\frac{1}{3}} \\
\displaystyle SH\Bigr|_{\substack{forced}}=\rho_{atm}C_P \;A\;u\; (T_{atm}-T_{surf})
\end{array}
\end{cases}
\label{eq_SH}
\end{equation}

with $k_{atm}$ the atmospheric thermal conductivity,  $C_P$ the specific heat,  $\nu$ the kinematic viscosity, $g$ the gravity, $A$ the wind function from Eq. (\ref{eq_funct_forced}), and $u$ the wind speed. Though the sensible heat losses are small with respect to the radiative losses for present-day surface pressure ($SH \sim$ 2.5-11 W/m$^2$ $\ll \epsilon \sigma T^4$, see \citet{schubert_2013}), they increase proportionally with the atmospheric density $\rho_{atm}$ so the atmospheric temperature $T_{atm}$ must be carefully related to $T_{surf}$ in order to avoid an overestimation of the convective losses at high surface pressure.

For present-day Marsand for the early Mars cases with low surface pressure (P$\leq$ 100 mbar), we used $b$=0.2 as in \cite{dundas_2010}. For the present-day test case, using $b$=0.2 produced a noontime temperature difference $T_{surf}-T_{atm}$~ $\sim$ 24K, which is similar to the difference between the ground temperature and the mast air temperature $\sim$ 20K measured by MSL in Gale crater (\cite{martinez_2017}).   For the high surface pressure cases (P$\geq$ 500 mbar), $b$ was tuned to 0.1 to account for the increased efficiency of the upward heat transport by turbulent eddies which tend to homogenize the air temperatures near the surface ($T_{surf}-T_{atm}$ decreases). The choice $b$=0.1 produces sensible heat fluxes at P= 1bar which are similar to Earth's convective losses: 24-83 W/m$^2$ in average, with peak afternoon values not exceeding $250-300$ W/m$^2$ \citep{schubert_2013,wang_2001}.

\subsubsection*{Conductive flux}

The conductive flux is $Q_{cond}=k\frac{\partial T}{\partial Z}$ with $k$ the thermal conductivity of regolith, $T$ the temperature and $Z$ the depth was solved using an implicit finite difference scheme with Von Neumann boundary conditions at the top and bottom of the domain.

\subsubsection*{Table of coefficients used for the time-marching surface energy balance}

\begin{table}[H]
\scriptsize
\centering
\begin{tabular}{|c|c|}
\hline
Parameter & Value  \\
\hline
\hline
Thermal conductivity $k$ & 0.065 $Wm^{-1}K^{-1}$  \\
Density regolith $\rho_{rego}$ &1500 $kg.m^{-3}$   \\
Spe. heat capacity regolith ${c_p}_{rego}$ & 750 $J.kg^{-1}K^{-1}$  \\
Molar mass CO$_2$ $M_{CO2}$ & 44 $g.{mol}^{-1}$   \\
Spe. heat capacity CO$_2$ ${c_p}_{CO2}$ & 770 $J.kg^{-1}K^{-1}$  \\
Thermal conductivity CO$_2$ $k_{atm}$ & 0.02 $Wm^{-1}K^{-1}$  \\
Wind $u$& 5  $ms^{-1}$  \\
albedo $a$ &  Varies as in Table \ref{table_OLR_IRD}   \\
Atm. parameter $b$ &  0.1-0.2   \\
emissivity $\epsilon$&  1 \\
obliquity $\theta$ &  25$^o$  \\
geothermal flux $F_{geo}$ &  0.03$Wm^{-2}$  \\
\hline
\end{tabular}
\caption{Simulation parameters and physical constants for Martian regolith. The values are close to the ones used in the Ames-GCM and consistent with \citet{Grott_2007};\citet{stoker_1991}; \citet{zent_2009}}
\label{soil_table}
\end{table}

\subsection*{E. Coefficients used for the ice thickness calculation}

\begin{table}[H]
\scriptsize
\centering
\begin{tabular}{|c|c|p{2cm}|}
\hline
Parameter & Value &Comment \\
\hline
\hline
Thermal conductivity $k_{ice}$ & 780/T+0.615$W/m/K$&    \citep{mckay_1985}\\
Density ice $\rho_{ice}$ & 900 $kg.m^{-3}$ & Assume pure ice \\
Latent heat $L$ &334000$J/kg$ &\\
Wind $u$& 5  $ms^{-1}$ &Same as Eq. (\ref{energy_budget_eq}) \\
Relative humidity $RH$ & 0\%&hyper-arid climate \\
Geothermal flux $F_{geo}$ &  0.03$Wm^{-2}$&Same as Eq. (\ref{energy_budget_eq})  \\
Albedo $a_{ice}$ & 0.66 & Same as Lake Untersee (\emph{McKay, private comm.})  \\
\% of energy <700 nm $f_{700}$ &  50\%  &\citep{mckay_2004} \\
\hline
\end{tabular}
\caption{Simulation parameters for the ice thickness calculation and physical constants for the ice}
\label{ice_table}
\end{table}

\subsection*{F. Analytical solution for the ice cover thickness}

Eq. (\ref{eq_ice_thick1}) is a non-linear equation in $\overline{Z}$. We re-organize it and define the constants $\alpha$ and $\beta$ as follows:

\begin{equation}
\begin{split}
& \overline{Z} =\frac{b \;ln(\frac{T_o}{\overline{T_R}}) +c \left(\overline{T_R}-T_o\right)-F_{sun} h \left(1-e^{\frac{-\overline{Z}}{h}}\right)}{F_{geo}+\rho_{ice} v L(1-\gamma_{\text{glacier}})} \\
&\Rightarrow \overline{Z}\left(F_{geo} +\rho_{ice} v L \left(1-\gamma_{\text{glacier}}\right)\right)= b \;ln(\frac{T_o}{\overline{T_R}}) +c \left(\overline{T_R}-T_o\right)\\
& \hspace{13em} - F_{sun} h +F_{sun} h e^{-\frac{\overline{Z}}{h}}\\
&\Rightarrow \overline{Z}\left(F_{geo} +\rho_{ice} v L \left(1-\gamma_{\text{glacier}}\right)\right) \\
&  + F_{sun} h - b \;ln(\frac{T_o}{\overline{T_R}}) -c \left(\overline{T_R}-T_o\right) =F_{sun} h e^{-\frac{\overline{Z}}{h}}\\
&\Rightarrow\frac{\overline{Z}}{h} \underbrace{\frac{F_{geo} +\rho_{ice} v L \left(1-\gamma_{\text{glacier}}\right)}{F_{sun}}}_{\alpha} \\
&\quad+ \underbrace{1- \frac{b \;ln(\frac{T_o}{\overline{T_R}}) +c \left(\overline{T_R}-T_o\right)}{F_{sun} h}}_{\beta}=e^{-\frac{\overline{Z}}{h}}\\
&\Rightarrow\frac{\overline{Z}}{h} \alpha + \beta=e^{-\frac{\overline{Z}}{h}}\\
\end{split}
\label{eq_mckay85_rearanged}
\end{equation}

We introduce a new variable $Y$ defined as :

\begin{equation}
\overline{Z}= \left(Y-\frac{\beta}{\alpha}\right)h
\label{eq_change_var}
\end{equation}

and use it to change variable in Eq. (\ref{eq_mckay85_rearanged}):

\begin{equation}
\begin{split}
&\frac{\overline{Z}}{h} \alpha + \beta=e^{-\frac{\overline{Z}}{h}}\\
&\Rightarrow \frac{\left(Y-\frac{\beta}{\alpha}\right)\cancel{h}}{\cancel{h}}\alpha +\beta= e^{-\frac{\left(Y-\frac{\beta}{\alpha}\right)\cancel{h}}{\cancel{h}}} \\
&\Rightarrow \alpha Y = e^{-Y} e^{\frac{\beta}{\alpha}} \\
&\Rightarrow  Y e^{Y} =  \frac{e^{\frac{\beta}{\alpha}}}{\alpha} \\
&\Rightarrow \underbrace{W\left( Y e^{Y}\right)}_{Y} =  W\left(\frac{e^{\frac{\beta}{\alpha}}}{\alpha}\right)
\end{split}
\label{eq_mckay85_rearanged2}
\end{equation}

Where $W$ is the Lambert-W function, defined as:
\begin{equation}
Y=W(Y e^Y)
\label{eq_lambert_property}
\end{equation}

Finally, from Eq. (\ref{eq_lambert_property}) and Eq. (\ref{eq_change_var}) we derive the analytic solution for the ice thickness in Eq.  (\ref{eq_analytic_mckay85_k_var}):

\begin{equation}
 \overline{Z}= h\left(W\left(\frac{e^{\frac{\beta}{\alpha}}}{\alpha}\right) -\frac{\beta}{\alpha}\right)\\
\end{equation}

with $\alpha$ and $\beta$ defined in Eq. (\ref{eq_mckay85_rearanged}). The Lambert-W function is available in many scientific programming languages and a uniform approximation from \citet{winitzki_2003} with a relative error $<$10$^{-2}$ is repeated here for convenience:

\begin{equation}
\begin{cases}
\begin{aligned}
W(x)&=\Bigg(2 \log(1+B \sqrt{2 e x+2}) \\
&-\log(1+C \log(1+D \sqrt{2 e x+2}))+E \Bigg)\\
& \times \frac{1}{1+1/(2 \log(1+B \sqrt{2 e x+2})+2A) }\\
&\\
& e=2.71828... \quad A = 2.344 \quad B = 0.8842 \\
& C = 0.9294 \quad D = 0.5106 \quad E =-1.213
\end{aligned}
\end{cases}
\label{eq_lambert_approx}
\end{equation}

\section*{Acknowledgements}

The authors thank Michael D. Smith for his valuable assistance to update the correlated-k coefficients used in the NASA Ames radiative transfer code, the NASA Ames Mars Climate Modeling Center (MCMC) who provided support this project as well as  the anonymous reviewer and Dr Edwin Kite for their careful reviews which greatly improved this paper.

\section*{References}


\bibliographystyle{apalike}

\bibliography{./paper_ice_rev2}
\end{document}